\documentclass[12pt,preprint]{aastex}

\def\lea{\mathrel{<\kern-1.0em\lower0.9ex\hbox{$\sim$}}}
\def\gea{\mathrel{>\kern-1.0em\lower0.9ex\hbox{$\sim$}}}

\slugcomment{Accepted for Publication in AJ}
\shorttitle{Star Clusters in M51} \shortauthors{Hwang \& Lee}

\begin{document}

\title{A Catalog of Bright Star Clusters in the Interacting Galaxy M51\altaffilmark{*}}

\author{\sc Narae \ Hwang and Myung Gyoon \ Lee}
\affil{Astronomy Program, Department of Physics and Astronomy,\\
Seoul National University, Seoul 151-747, Korea}
\email{nhwang@astro.snu.ac.kr, mglee@astrog.snu.ac.kr}

\altaffiltext{*}{Based on observations made with the NASA/ESA {\it
Hubble Space Telescope}, obtained from the data archive at the
Space Telescope Science Institute, which is operated by the
Association of Universities for Research in Astronomy, Inc., under
NASA contract NAS 5-26555.}

\begin{abstract}
We present a catalog of star clusters with
$V_{F555W}<23$ mag detected in the interacting spiral galaxy M51
system based on the mosaic images taken with $HST$ ACS by the
Hubble Heritage Team. We have selected about 3,600 clusters based
on their morphological information through the visual inspection.
The final star cluster catalog
includes 2,224 clusters that are relatively well isolated and have
a circular shape. The star clusters in M51 are mostly distributed
around the spiral arms of NGC 5194. The color-magnitude diagrams
show that most of the star clusters in M51 are bluer than
$(B_{F435W}-V_{F555W}) = 0.5$ and $(V_{F555W}-I_{F814W}) = 0.8$.
There are also some red star clusters with $(B_{F435W}-V_{F555W})
> 0.7$, uniformly distributed over the M51 field. Some of these red
clusters are suspected to be a part of the halo or old disk population
based on their old ages ($t \geq 10^9$ yrs) and their spatial
distribution. The luminosity function of the star clusters is
fit well by a single power law 
with $\alpha = -2.59 \pm 0.03$ for the range $-10.0 <M_V < -8.0$ mag.
We find that the size distribution of the star clusters can be fit with
three Gaussian components with peaks at effective radii of 2.27, 4.80 and 7.51 pc.
Some large star clusters with red color are faint
fuzzy clusters, and they are distributed not only around NGC 5195
but also around NGC 5194. These faint fuzzies are found to display
an elongated spatial distribution, while the normal compact red
clusters show a relatively uniform distribution around NGC 5194.

\end{abstract}


\section{Introduction}

Studying star clusters is important for understanding the star
formation history of their host galaxies and the star cluster population
itself. Globular clusters found in giant elliptical galaxies show
that most elliptical galaxies are very old systems that ceased to
form stars or star clusters a long time ago (see Lee 2003 and
references therein). On the contrary, many star clusters in
late-type galaxies display rather blue colors, indicating very
recent star formation activities in the corresponding galaxies
\citep{lee06}. For interacting or merging galaxies, it is known
that there are some very bright and massive star clusters such as
super-star clusters (SSCs) in Antennae galaxy \citep{whi99}. These
massive clusters are found in galaxies that are undergoing an
active star formation probably induced by dynamical interactions.

The existence of such massive star clusters indicates that burst-like star formation
activities in some interacting or merging galaxies are intimately
correlated with the dynamical events undergone by their host
galaxies. This also suggests that, with regard to normal galaxies,
star clusters can also be used as an excellent tracer of star
formation activity in interacting or merging galaxies. However,
finding star clusters in galaxies is not easy. This is possible
only for nearby galaxies even after the advent of the {\it Hubble
Space Telescope} (HST), which provides a superb spatial resolution
and photometric depth. M51 is located at a distance of 8.4 Mpc
\citep{fel97} so that it is one of the most suitable
interacting galaxies for star cluster survey using the $HST$.

M51 is an interacting galaxy system that is composed of a face-on
Sbc-type spiral galaxy NGC 5194 and a fainter companion, the SB0
galaxy NGC 5195. It is abundant with interstellar media
(see a recent study by \citet{schu07}). Several conditions
with M51 make it suitable for the star cluster study using the $
HST$. (1) It is possible to marginally resolve the star clusters
at the distance of M51 with the resolving power of the $HST$. (2)
M51 is in a nearly face-on configuration, which allows a direct
view of the entire disk structure. (3) There are some previous
theoretical studies \citep{too72,sal00a,sal00b} suggesting that
these two galaxies may have experienced a single or multiple
encounters some hundreds of Myrs ago. That is, the properties and
epoch of the dynamical events are known to some extent, which
provides the base timeline for comparison with the derived star
formation history.

There are several previous studies on the star cluster population
in M51 that have used $HST$ data. In \citet{bik03} and
\citet{bas05a}, approximately 1,000 star clusters were detected in
a small region near the M51 center based on $HST$ WFPC2 data.
\citet{bas05a} argued that the cluster formation rate seems to
increase considerably around the expected time of the second
encounter according to the derived age distribution of the star
clusters. It was also suggested that young and massive star clusters in
M51 tend to form in complexes rather than in isolation
\citep{bas05b} and even clusters as massive as $10^6 M_{\odot}$
can be disrupted in the M51 disk within a few Gyrs after their
formation \citep{gie05}.

On the other hand, \citet{lee05} adopted a conservative policy for
star cluster selection and reported only about 400 resolved star clusters
based on $HST$ WFPC2 data that covered the larger area than used in the earlier studies.
They showed that the
age distribution of these resolved clusters displays a broad peak
at several hundred Myrs, and this is consistent with the epoch of
the first encounter predicted by the theoretical models. Another
important point noted in \citet{lee05} is that the star cluster
sample can be so easily contaminated that star clusters should be
selected with great care for the case of M51, since they are only
marginally resolved and can be easily confused with other sources
such as bright point sources.

Previous studies on the star clusters in M51 have a common limit
regarding the $HST$ data used, which is the incomplete areal
coverage of M51. Since the existing $HST$ WFPC2 observation of M51
only covered the central parts of NGC 5194 and NGC 5195 along with
some parts of the spiral arms, the star cluster sample reported in
the previous studies can be biased toward young and massive
clusters mostly found around the galaxy center and the spiral
arms. However, in order to investigate the complete star formation
history using the star clusters in M51, we need to carry out a
star cluster survey over the field of entire galaxy.
Although the globular cluster search in M51 performed by
\citet{cha04} returned 34 globular cluster candidates, it still
suffers from an incomplete areal coverage.

We have carried out a star cluster survey over the
M51 field using the newly released Hubble Heritage Data that
covers the main body of NGC 5194.
As a first step to the study of the star cluster
population in M51, we have searched for both old (red) and young
(blue) star cluster candidates based on a homogeneous detection
and classification policy that employs the visual inspection of
sources before the final registration of the star cluster
candidates. This method does not guarantee the complete census of
the star clusters due to the inherent limitation of manual
inspection. However, this approach can be more efficient to lower
the possible contamination in star cluster samples, as shown by
\citet{lee05}. Some preliminary results of this study showed the
overabundance of faint fuzzy clusters around NGC 5195
\citep{hwa06} and implied a possible correlation between the
dynamical encounters of host galaxies and age distribution of star
cluster candidates \citep{hwa07a}.

In this paper, we present the results of M51 star cluster survey,
including the methodology of finding and classifying star cluster
candidates. We also present the resulting final catalog of star
cluster candidates with $V_{F555W}<23$ mag. This cluster catalog
has served as the master catalog in \citet{hwa06,hwa07a}. Then, we
investigate some photometric properties and the spatial
distribution of these cluster candidates. Finally, some properties
of large clusters such as faint fuzzy clusters found in the M51
system are discussed. During this study, there came a
report by \citet{sch07} that identified about 1300
star cluster candidates in M51 based on the same Hubble Heritage
Data. We present an independent and complementary study that was
carried out by using a different approach. We will discuss some
common characteristics of star clusters shown in both studies at
the end of this paper.

\section{Observation and Data}

The data used in this study were obtained by the Hubble Heritage
Team to commemorate the $HST$'s 15th anniversary using the $HST$
ACS with F435W, F555W, F814W, and F658N filters through $HST$
observing program 10452 (P.I.: Steven V. W. Beckwith). The
observed field is as large as about $6.8' \times 10.5'$ centered
on NGC 5194. The accumulated exposure times are 2,720 s for the
F435W band and 1,360 s for the F555W and F814W bands. All the
necessary data reduction processes were done by the STScI
including multi-drizzling and image combination before the data
release. More detailed information on the observation and data
reduction is given in \citet{mut05}. \footnote{See also
http://archive.stsci.edu/prepds/m51/.}

We adopted a distance of $8.4\pm0.6$ Mpc ($(m-M)_0=29.62$) to M51
determined from the planetary nebula luminosity function in M51
\citep{fel97}. The corresponding linear scale is 40.7 parsecs per
arcsecond. Since the $HST$ ACS mosaic data have pixel sizes of
$0.05\arcsec$, the resultant projected physical scale covered by
one pixel is approximately $2.04$ pc. The foreground reddening
toward M51 is low, $E(B-V)=0.035$, and the corresponding
extinctions are $A_B=0.150$, $A_V=0.115$, and $A_I=0.067$ mag
\citep{sch98}. The total magnitudes and colors are
$B^T=8.96\pm0.06$ mag and $(B^T-V^T)=0.60\pm0.01$ for NGC 5194,
and $B^T=10.45\pm0.07$ mag and $(B^T-V^T)=0.90\pm0.01$ for NGC
5195 \citep{dev91}. At the adopted distance (without internal
extinction correction), the absolute magnitudes are $M_B^T=-20.81$
mag, $M_V^T=-21.38$ mag for NGC 5194, and $M_B^T=-19.32$ mag and
$M_V^T=-20.19$ mag for NGC 5195.

\section{Star Cluster Detection and Classification}

\subsection{Source Detection and Photometry}
\label{phot}

Detecting a source is a demanding task if the source is
embedded in a varying background, such as the spiral arms of M51.
To work around this difficulty, a representative background
map was constructed by median filtering of the F555W band image of
M51. The $RMEDIAN$ task of $IRAF$ package was employed for this
process by setting the inner and outer radii of the filtering ring
to 20 and 30 pixels, respectively. Then, the constructed
background map was subtracted from the original F555W band image
of M51 to produce a detection image in which only the resolved
sources remain. Source detection was done by running SExtractor
\citep{ber96} on this detection image. A detection threshold of
$4\sigma$ and minimum contiguous detected area of five pixels were
used for finding sources in the detection image. The total number
of detected sources is approximately 120,000.

The flux of the detected objects was measured using SExtractor run
in the dual mode in the original F435W, F555W, and F814W band
images. Among the various photometric measurement values produced
by SExtractor, we adopted a flux measured using an aperture with
$r = 6$ pixels ($0.3\arcsec \simeq 12$ pc), which is about three times the
average FWHM of the point sources in the images, as the
representative flux of an object. The colors of the objects were
also calculated based on the photometric magnitudes measured with
the same $r=6$ pixel apertures in the corresponding bands. The
instrumental magnitudes in the F435W, F555W, and F814W bands were
calibrated using the photometric zero points of the Vega magnitude
system for the WFC of $HST$ ACS provided by \citet{sir05}.
Hereafter, we use $B$, $V$, and $I$ to denote $B_{F435W}$,
$V_{F555W}$, and $I_{F814W}$, respectively.

\subsection{Photometry Comparison with Previous Studies}

There are two sets of photometry data for the M51 star cluster
candidates available in the literature: one from \citet{bik03} and the other from
\citet{lee05}. Since \citet{lee05} have already shown that their
photometry measurements are in good agreement with those in
\citet{bik03}, we used the more recent photometry data from
\citet{lee05} for comparison purposes. In \citet{lee05}, aperture
photometry was derived using the aperture of $r = 3$ pixels and
was corrected for the finite aperture effect by adding $-0.39$ mag for
PC and $-0.31$ mag for WF. The aperture size of $r=3$ pixels in WF
of WFPC2 corresponds to 6 pixels ($0.3 \arcsec$) in ACS, which is
our adopted aperture size for photometry. We compared our
photometry data with that of WF CCDs in \citet{lee05} after
removing the aperture corrections.

\begin{figure}
\plotone{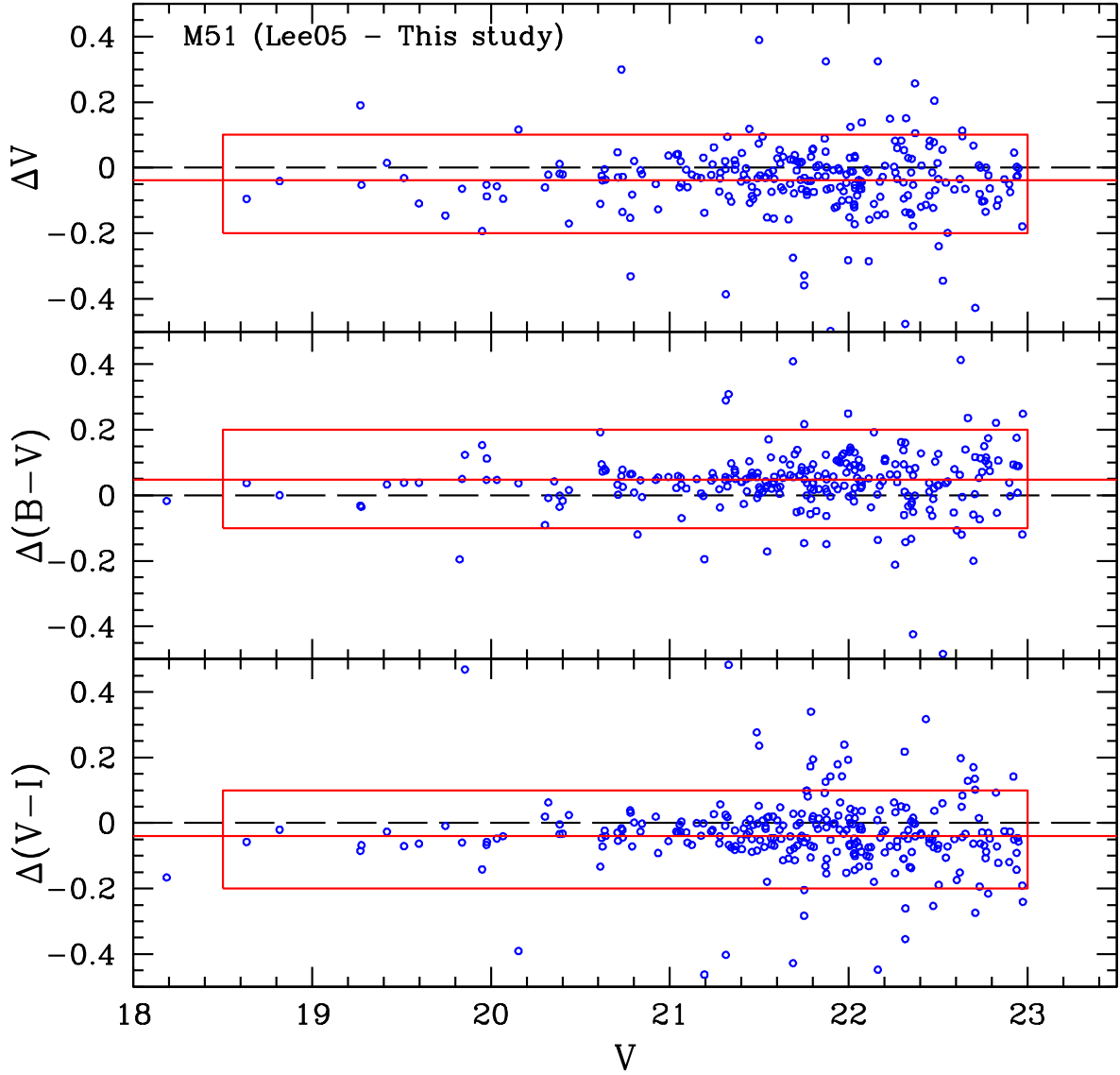} \caption{ Comparison of photometry of M51
clusters. Boxes represent the boundary for the data used to calculate
the mean offsets that are marked by the sloid lines.
\label{leecomp}}
\end{figure}

Figure \ref{leecomp} shows the result of the photometry
comparison. Our photometric measurements are in reasonably good
agreement with those of \citet{lee05}. The mean differences in
photometry for about 230 common objects are $\Delta V = -0.039 \pm
0.063$ mag, $\Delta (B-V) = 0.048 \pm 0.055$ mag, and $\Delta
(V-I) = -0.040 \pm 0.057$ mag, where $\Delta$ means \citet{lee05}
minus this study, and the errors are standard errors of the mean.

\subsection{Star Cluster Classification}
\label{visual}

At the distance of M51, typical star clusters appear to be
somewhat more extended than stars in the $HST$ images (see Figure
2 of \citet{lee05}). Therefore, it is possible to select cluster
candidates using the morphological parameters provided by
SExtractor. We used the cluster sample of \citet{lee05} as the
training set for constraining the parameter spaces of the objects'
FWHM, ellipticity (hereafter `$e$'), and stellarity in which most
of the star clusters would be found. This test revealed that most
of the star clusters in \citet{lee05} are found to belong to two
groups in the parameter spaces of FWHM, $e$, and stellarity:
Group 1 with low stellarity $< 0.5$, $2.4 < FWHM < 20$ pixels, and $e < 0.6$,
and Group 2 with high stellarity $> 0.9$, $2.4 < FWHM < 40 $ pixels,
and $e < 0.6$. The number of selected candidates
with $V < 23$ mag is 7,300 (about 5,900 for Group 1 and about
1,400 for Group 2). For a comparative reference, about 1,100 point
sources with $V < 23$ mag, $1.0<FWHM<2.3$ pixels, $e<0.2$, and
stellarity $>0.9$ are also selected. Therefore, about 8,400
out of the approximately 120,000 detected sources are
selected for the final visual classification.

The star cluster candidates in M51 can be found in various shapes
and environments. Therefore, during the visual classification, we
considered the following two factors: (1) the width and curvature
of the radial profiles and (2) the shape of the 2D contour and
environmental effects. The width and curvature of the radial
profiles are the most basic parameters used to select the star
cluster candidates. These criteria mainly serve to distinguish
star cluster candidates from point sources. The FWHM of the star
cluster candidates in M51 is about 3 pixels or larger, while the
FWHM of the point sources is about $2.1 - 2.3$ pixels in $HST$ ACS
data. The large FWHM of the star cluster candidates is also
reflected in the curvature of the radial profiles. The radial
profiles of the point sources show a very steep decrease away from
the core and they reach the background level within a radius of
2.5 pixels. On the other hand, the radial profiles of the star
cluster candidates have a relatively gentle slope and extend to a
radius of more than 3 pixels.

The shape of the 2D contour reveals how circular a star cluster
candidate appears and how many subcomponents it has as.
We separate a sample of star cluster
candidates with circular shapes from another sample that shows
distorted contours and/or multiple peaks in a single contour.
Another important information that can be derived from the 2D
contour map is how many neighbors there are near the candidate.
That is, we can investigate if a star cluster candidate is
surrounded by several sources that are clustered together or
whether a candidate has distorted contours or many subcomponents,
indicating the existence of very nearby unresolved sources or
multiple sources that are merged together. This helps to separate
well isolated star clusters with circular shape from those with
elongated or irregular shape and/or any prominent neighbors.

\begin{figure}
 \begin{center}
\includegraphics[scale=0.65]{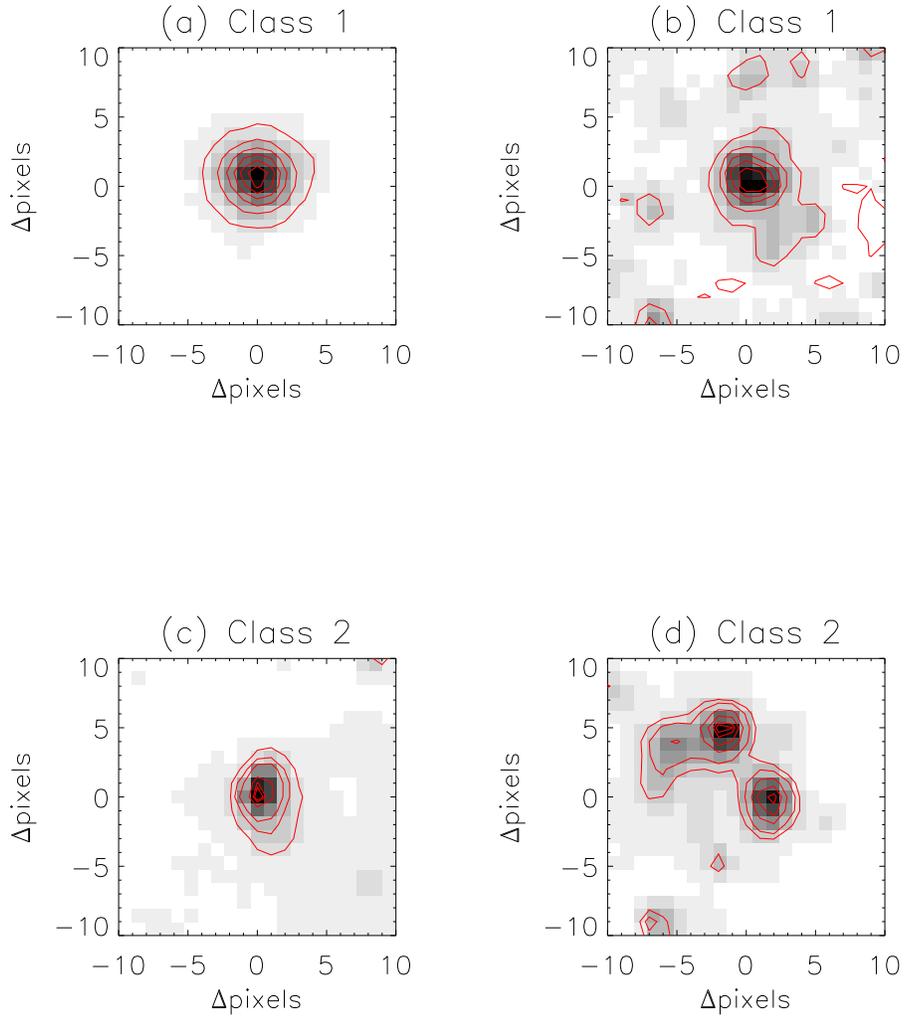}
 \caption{Grayscale maps of the representative $V$-band images of a few cluster candidates
overlaid with intensity contours. Panels (a) and (b) show Class 1 cluster candidates, while
panels (c) and (d) show Class 2 cluster candidates. See the text for details.
\label{contour}}
 \end{center}
\end{figure}

>From the visual classification based on these criteria, we define
a `Class 1' cluster sample as `cluster candidates' that have a very
well-defined nearly circular shape, and no prominent nearby neighbors.
Another class of star cluster candidates defined is a `Class 2'
cluster sample that includes `cluster candidates' with an elongated shape as 
well as irregular structures and/or multiple neighbors.
Figure \ref{contour} shows 2D contour maps of a few Class 1 and Class 2
cluster samples. After the visual
inspection of about 8,400 candidate extended candidates and point
sources, we have selected 2,224 Class 1 and 1,388 Class 2 star
cluster candidates.\footnote{Hereafter, Class 1 and Class 2 star
cluster candidates will be referred to as just `star clusters'
even though they are not yet spectroscopically confirmed to be genuine
star clusters.} It turns out that about $50\%$ of Group 1
candidates are classified as Class 1 ($\sim 35\%$) or Class 2
($\sim 15\%$) clusters, while about $23\%$ of Group 2 are
classified as Class 1 ($\sim 16\%$) or Class 2 ($\sim 7\%$)
clusters.

\subsection{Aperture Correction}
\label{apc}

\begin{figure}
\plotone{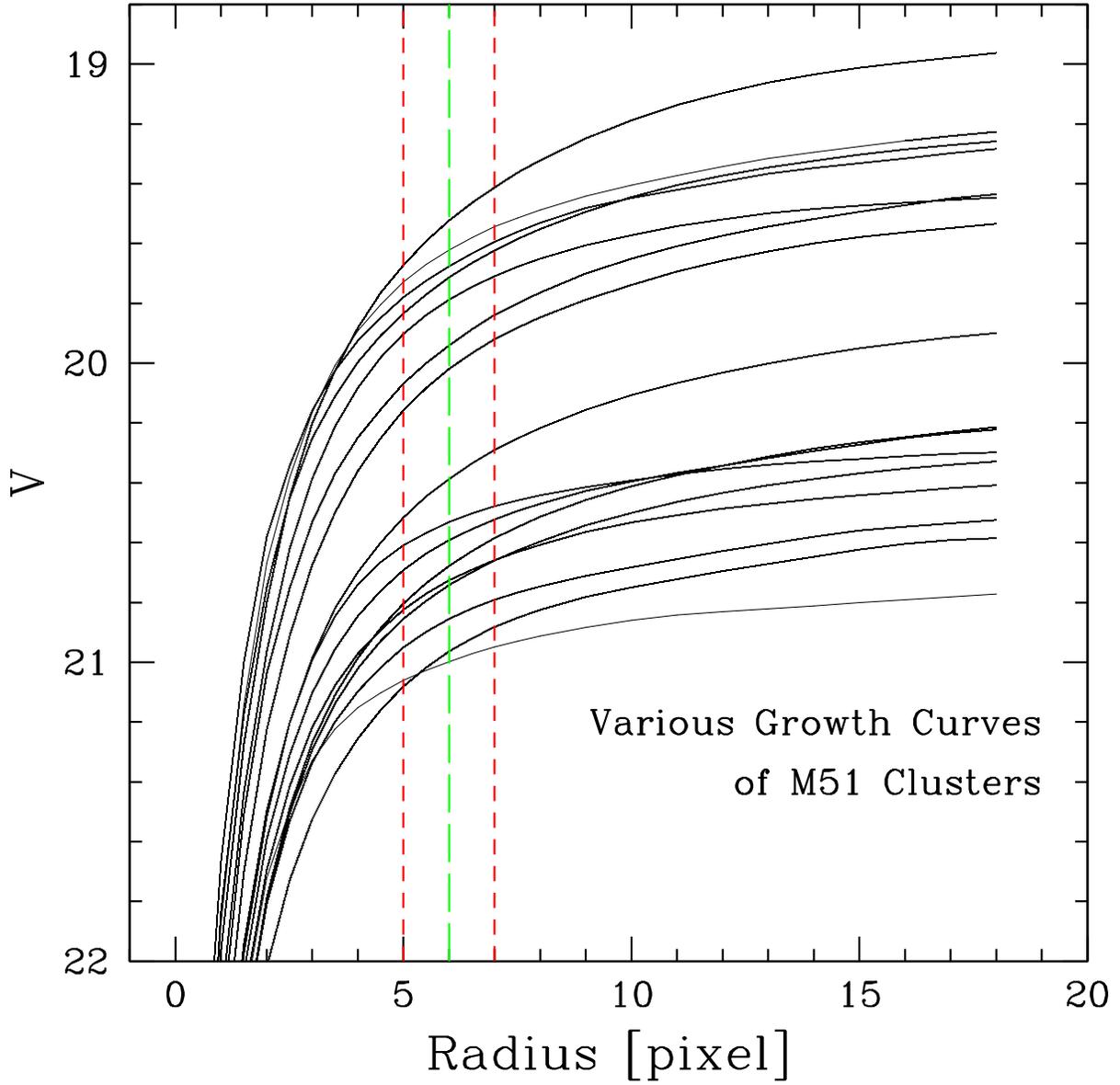}
\caption{
 Integrated magnitude profiles of isolated bright M51 star clusters.
 A long dashed line marks the size of the aperture we adopted for our aperture
 photometry of star clusters. Note that the magnitude differences (and the slope)
 between at $r=5$ pixels and at $r=7$ pixels (marked by short dashed
 lines) vary from one cluster to another.
 \label{apc1}}
\end{figure}

The aperture photometry of sources requires the correction for the
light lost due to the finite size of the aperture used. For point
sources such as stars, the amount of aperture correction is
determined by the instrumental characteristics of $HST$.
Therefore, we can apply a single value of aperture correction to
the photometry data of point sources. However, for extended
sources such as star clusters in M51, a single value of correction
cannot be defined based on a certain size of the aperture since
the luminosity profile of one cluster varies depending on its size and compactness.
Figure \ref{apc1} shows the integrated magnitude profiles of 16
star clusters in M51, which are found in well-isolated locations
and have nearly circular shapes. The effective radii $R_{\rm eff}$ of
these clusters range from 2.85 pc to 8.19 pc with a median value
of 5.01 pc. It is evident that the shape of the profile depends on
the size and/or the compactness of a cluster, which varies from
one cluster to another. This leads to a different amount of
aperture correction for each star cluster.

\begin{figure}
\plotone{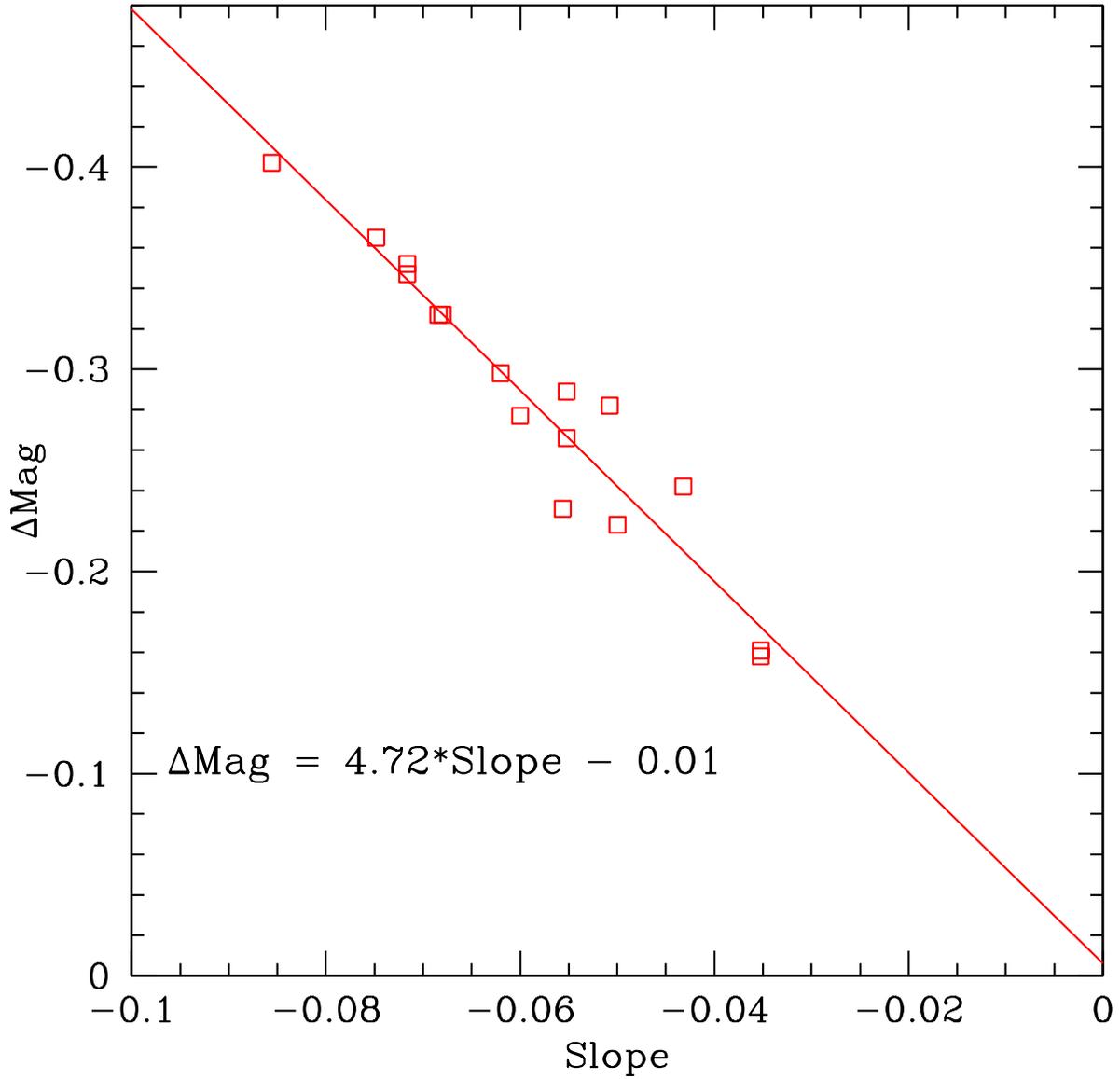} \caption{
 Relationship between the necessary aperture correction $\Delta Mag$ and
 the slope of the integrated magnitude profiles defined at $r=6$ pixels
 for the same isolated star clusters, as used in Fig. \ref{apc1}.
 The solid line shows the linear fit relation between the slope
 and the aperture corrections.
\label{apc2}}
\end{figure}

We investigated the correlation between the slope of the
integrated magnitude profile around the fixed radius of $r = 6$
pixels, the same size as that of the aperture used for the
photometry, and the amount of necessary correction calculated to
the outer radius of $r = 18$ pixels for each star cluster. The
slope of the integrated magnitude profile for each cluster at
$r=6$ pixels was derived from the difference between the $V$-band
integrated magnitudes for $r=5$ and $7$ pixels. As shown in Figure
\ref{apc2}, there is a good correlation between the slope of the
profile and the amount of magnitude correction $\Delta Mag$. The
relationship between the slope and $\Delta Mag$ is derived by
linear fitting: $\Delta Mag = 4.72 \times slope - 0.01$ for
$-0.09< slope < -0.03$ with rms $= 0.02$. However, to prevent the
overestimation of the aperture correction, we restricted the range
of the aperture correction to $-0.48< \Delta Mag < 0.0$ (and $-0.1
< slope < 0$). The aperture-corrected magnitude is used for
deriving the total magnitude in the $V$ band.

\subsection{Size Measurement}
\label{size}

We measured the effective radii of the selected clusters using
ISHAPE provided by \citet{lar99}. We used the MOFFAT15 model
profile for model fit since \citet{lar99} reports that the
MOFFAT15 model effectively reproduces the intrinsic half-light
radii of the input model star clusters with various model
profiles. The model fit was tried to $r = 8$ pixels from the
center of the star clusters. The goodness of fit was checked by
using the CHISQR value as well as the residual and subtracted
images produced by ISHAPE. The resultant FWHM values of the star
clusters were transformed into the half-light radii by multiplying
by 1.13, according to the guide in the ISHAPE manual.

We derived the empirical point spread function (PSF) necessary for
the ISHAPE run by carefully selecting the isolated stars scattered
evenly in the neighborhood of NGC 5195. The field around NGC 5194
is more densely populated with compact star clusters and it is
difficult to avoid contamination by starlike clusters during the
PSF candidate star selection. The field of NGC 5195, on the other
hand, is not crowded. The possible PSF dependence on the different
chips of $HST$ ACS is ignored since NGC 5195 field is already
covered by two independent $HST$ ACS pointings where our PSF
candidate stars are uniformly distributed.

To test the dependence of the size measurements on the selection
of the PSF candidate stars, we ran another ISHAPE operation on
about 200 bright star clusters with $V<21$ mag using a new
empirical PSF generated from about 20 stars found around the
spiral arms of NGC 5194. We found that the difference in the
measured sizes of the star clusters is $\Delta R_{\rm eff} < \pm 1$ pc
for most clusters: it is found that $\Delta R_{\rm eff}$ does not
exceed 3 pc. This result shows that the rate of difference in the
measured size is not greater than 20\% for most of the star
clusters, and this rate tends to be smaller as the size of the
clusters gets larger. For 115 star clusters with $R_{\rm eff} > 2$ pc,
which is about 1 pixel wide, the rate of star clusters with
$\Delta R_{\rm eff} < 20\%$ is approximately $0.80$.

\section{Star Cluster Catalog}
\label{cat}

We present a catalog of Class 1 star clusters with $V < 23$ mag.
Table \ref{gcat} lists a sample catalog of Class 1 star clusters
for a reader's guide and the full catalog will be available
electronically from the online Journal. Brief descriptions on the
data columns of the catalog are given as follows:

Column 1 lists an identification number of a star cluster.

Columns 2 and 3 list the J2000.0 right ascension (RA) and
declination (DEC) of a star cluster in degrees.

Columns 4 and 5 list the $V$-band magnitude and corresponding
error of a star cluster, measured with an aperture of $r = 6$
pixels.

Columns 6 and 7 list the $(B-V)$ color and corresponding error of
a star cluster.

Columns 8 and 9 list the $(V-I)$ color and corresponding error of
a star cluster.

Column 10 lists the extraction flag of a star cluster given by
SExtractor. Flag = 0 means that the star cluster is well isolated.
On the other hand, flag = 1 indicates that the cluster has
neighbors and flag = 2 indicates that the cluster is originally
blended with another source but deblended by SExtractor. If a star
cluster has flag = 3, it means that the cluster has a neighbor and
is blended with it but deblended by SExtractor.

Column 11 lists the stellarity index of a star cluster. This index
has a value between 1 (point sources) and 0 (extended sources).

Column 12 lists the FWHM of a star cluster in arcseconds
calculated under the assumption that the cluster has a Gaussian
profile.

Column 13 lists the effective radius or half-light radius of a
star cluster in pc measured using ISHAPE. See Section \ref{size}
for details.

Column 14 lists the ratio of $\chi^2$ defined as
$(\chi^2/\chi^2_0)$ calculated by ISHAPE during the effective
radius measurement. The $\chi^2$ value is the minimum $\chi^2$
calculated after the broadening of the input PSF, while $\chi^2_0$
is calculated using the input PSF without any broadening. This
ratio indicates how significant the deviation from the input PSF
is. The larger this value, the higher the significance of
deviation.

Column 15 lists the ellipticity of a star cluster calculated by
SExtractor using second-order image moments.

Column 16 lists the aperture-corrected $V$-band magnitude of a
star cluster following the method illustrated in Section
\ref{apc}.

Column 17 lists any optional comment for a star cluster. If the
cluster is classified as a faint fuzzy cluster, then a comment of
`FF' is added to this column.

\section{Photometric Properties of Star Clusters}

\begin{figure}
\plotone{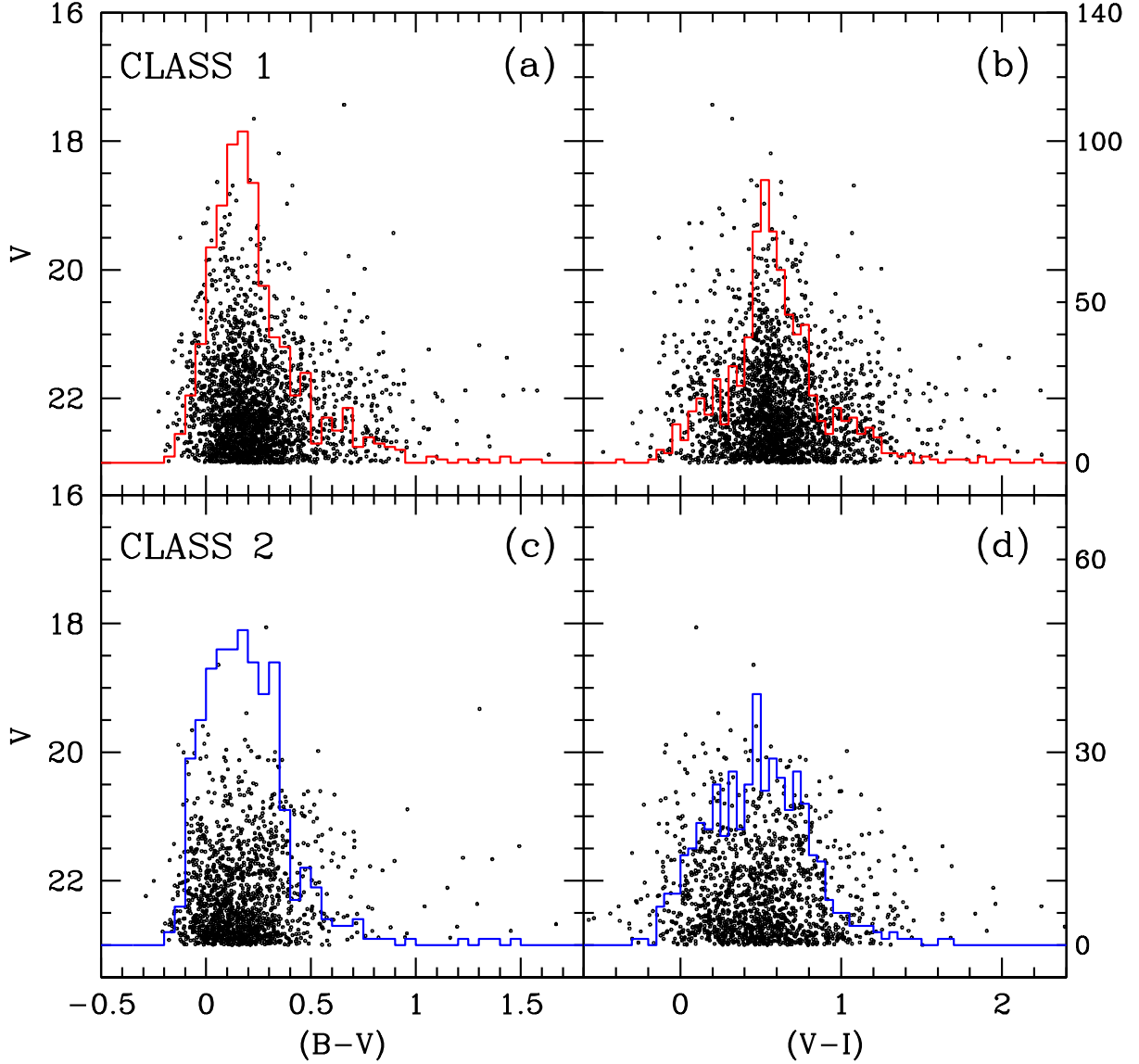} \caption{
 $V$ - $(B-V)$ and $V$ - $(V-I)$ color-magnitude diagrams of all star
 clusters with $V<23$ mag in M51. The overlaid $(B-V)$ and $(V-I)$ color histograms
 (corresponding y-axes are drawn in the right axis) 
 are for the star clusters with $V<22$ mag. The upper panels (a) and (b) are
 for Class 1 star clusters, while the lower panels (c) and (d) are for Class 2
 star clusters. The typical photometric errors in $(B-V)$ and $(V-I)$ at $V<23$ mag
 do not exceed $0.03$ mag in most cases.
\label{cmd}}
\end{figure}

Figure \ref{cmd} shows the $V$ - $(B-V)$ and $V$ - $(V-I)$
color-magnitude diagrams (CMD) of the star clusters in M51. In
these diagrams, it can be seen that M51 star clusters, regardless
of their classification, are found to be in the range of $-0.2
\leq (B-V) \leq 1.5$ and $-0.2 \leq (V-I) \leq 2.0$. The brightest
star cluster appears to be as bright as $V \approx 17.5$ ($M_V
\approx -12.1$) mag, which is even brighter than $\omega$Cen ($M_V
= -10.29$ mag) in our Galaxy. However, most Class 1 and Class 2
star clusters in M51 are blue with $(B-V)<0.5$ and $(V-I)<0.8$ and
fainter than $V \approx 20$ mag. This suggests that most star
clusters in M51 are young. There are some star clusters redder
than $(B-V) = 0.5$ and/or $(V-I) = 0.8$. The color spread of Class
1 star clusters is slightly larger than that of Class 2 star
clusters. That is, these red star clusters are more abundant in
Class 1 than in Class 2.

\begin{figure}
 \plotone{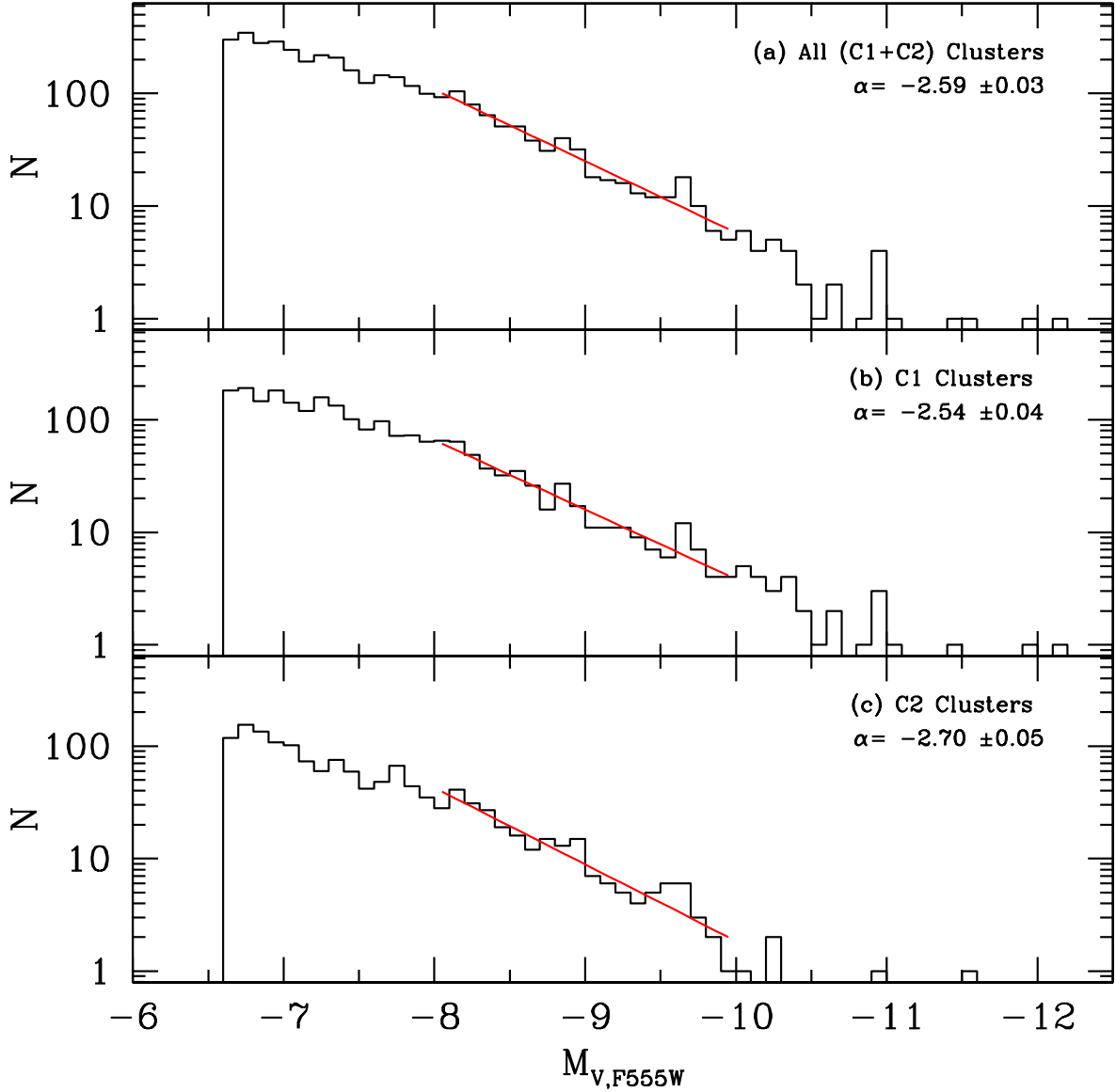} \caption{
  Luminosity functions of all (Class 1 and 2) clusters (panel (a)), Class 1 clusters (panel (b)),
  and Class 2 clusters (panel (c)). The faint end cut at $M_V = -6.6$ is caused
  by the criterion of $V<23$ adopted for the star clusters in this study.
  The solid line in each panel indicates the power-law fit of the luminosity function.
 The logarithmic slope $\alpha$ is derived through
  the fit over the range of $-10.0 < M_V <-8.0$.
\label{clf}}
\end{figure}

We derived the $V$-band luminosity function of star clusters,
displaying it in Figure \ref{clf}.
Figure \ref{clf} shows that the luminosity functions of star clusters
in M51, regardless of their classifications, are clearly of
power-law form and keep rising up to $M_V \sim -7.0$ mag.
The luminosity functions of star clusters become to be slightly flatter for 
the fainter magnitudes $M_V \ge -8.0$ mag, 
indicating that our cluster detection may be incomplete in this range. 
We restricted the magnitude range to $-10.0 <M_V < -8.0$ mag
where incompleteness of cluster detection is negligible, 
for fitting the data with a power law, $N dL \varpropto L^{\alpha} dL$.
It is found that the logarithmic slope $\alpha$ is $-2.59 \pm 0.03$ for all 
star clusters in M51 (Fig. \ref{clf}a), while
$\alpha = -2.54 \pm 0.04$ for Class 1 clusters (Fig. \ref{clf}b) and
$\alpha = -2.70 \pm 0.05$ for Class 2 clusters (Fig. \ref{clf}c),
showing only a little difference between the two cluster classes.
These slopes of luminosity functions are in good agreement with 
the value \citet{gie06b} derived from the WFPC2 data for M51 star clusters
for the range of $-11 < M_V < -8$ (see their Figure 7), $\alpha = -2.5 \pm 0.1$.
They are also similar to the value for NGC 6946 \citet{lar02} derived from the
WFPC2 data  for the range of $-10.75 < M_V < -8.0$, $\alpha = -2.46 \pm 0.12$.

On the other hand, \citet{gie06a} reported that there is a break at $M_V \approx -9.0$, 
in the luminosity function of M51 star clusters derived from the same ACS data as used in this
study. From fitting with a double power law they derived the slopes, 
$-1.93\pm0.03$ for $-6.32>M_V >-8.93$, and 
$-2.75\pm0.14$ for $M_V < -8.93$.
However, this break is not seen in the luminosity function derived in this study.
We calculated $\chi^2$ values for two cases:
one with a single power law using the same parameters derived in this study,
and the other with a double power law using the same parameters given by \citet{gie06a}.
We obtained $\chi^2 = 0.32$ for the double power law fit, which is about twice
larger than $\chi^2 = 0.15$ derived from the single power law fit,
suggesting that the single power law fit gives or a better fit.

\begin{figure}
\plotone{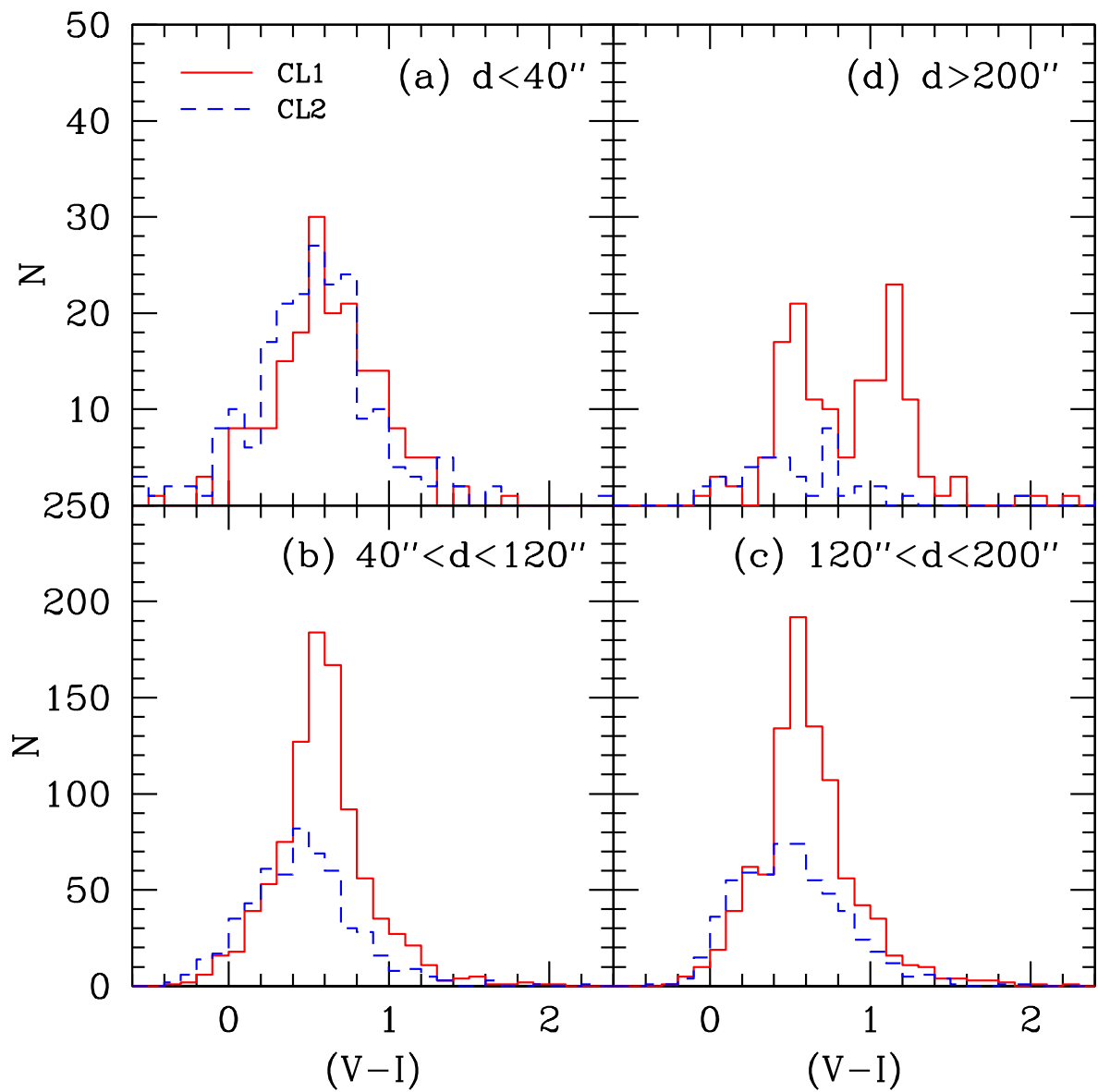} \caption{
 $(V-I)$ color histograms of M51 star clusters with $V<23$ mag in each annulus
 defined as $d<40\arcsec$ (panel (a)), $40\arcsec<d<120\arcsec$ (panel (b)),
 $120\arcsec<d<200\arcsec$ (panel (c)), and $d>200\arcsec$ (panel (d)), respectively.
 The color distribution of Class 1 star clusters is shown by solid lines,
 while that of Class 2 star clusters is shown by dashed lines.
\label{reghist}}
\end{figure}

We have investigated any radial dependence of the star clusters'
$(V-I)$ color distribution by dividing the observed M51 field into
four concentric annuli centered on the NGC 5194 center. Figure
\ref{reghist} shows that there is a distinct population of red
star clusters at a distance of more than $d = 200 \arcsec$ from
the center of NGC 5194, making the $(V-I)$ color distribution for
Class 1 clusters clearly bimodal in this region with peaks at
$(V-I) \sim 0.5$ and $\sim 1.2$. This indicates that Class 1
clusters in this outer region are actually composed of two
populations: a blue disk population and a red halo/old disk
population, which is also noted in Section \ref{spatdist}. On the
other hand, the $(V-I)$ color distributions of the star clusters
in the other three annuli display a color peak at $(V-I) \approx
0.5$ and appear to be well represented by a single Gaussian
distribution. The number ratio of Class 2 to Class 1 star clusters
is about 0.65 on an average for $d < 200 \arcsec$, but it
decreases to about 0.28 for $d > 200 \arcsec$.

\begin{figure}
\plotone{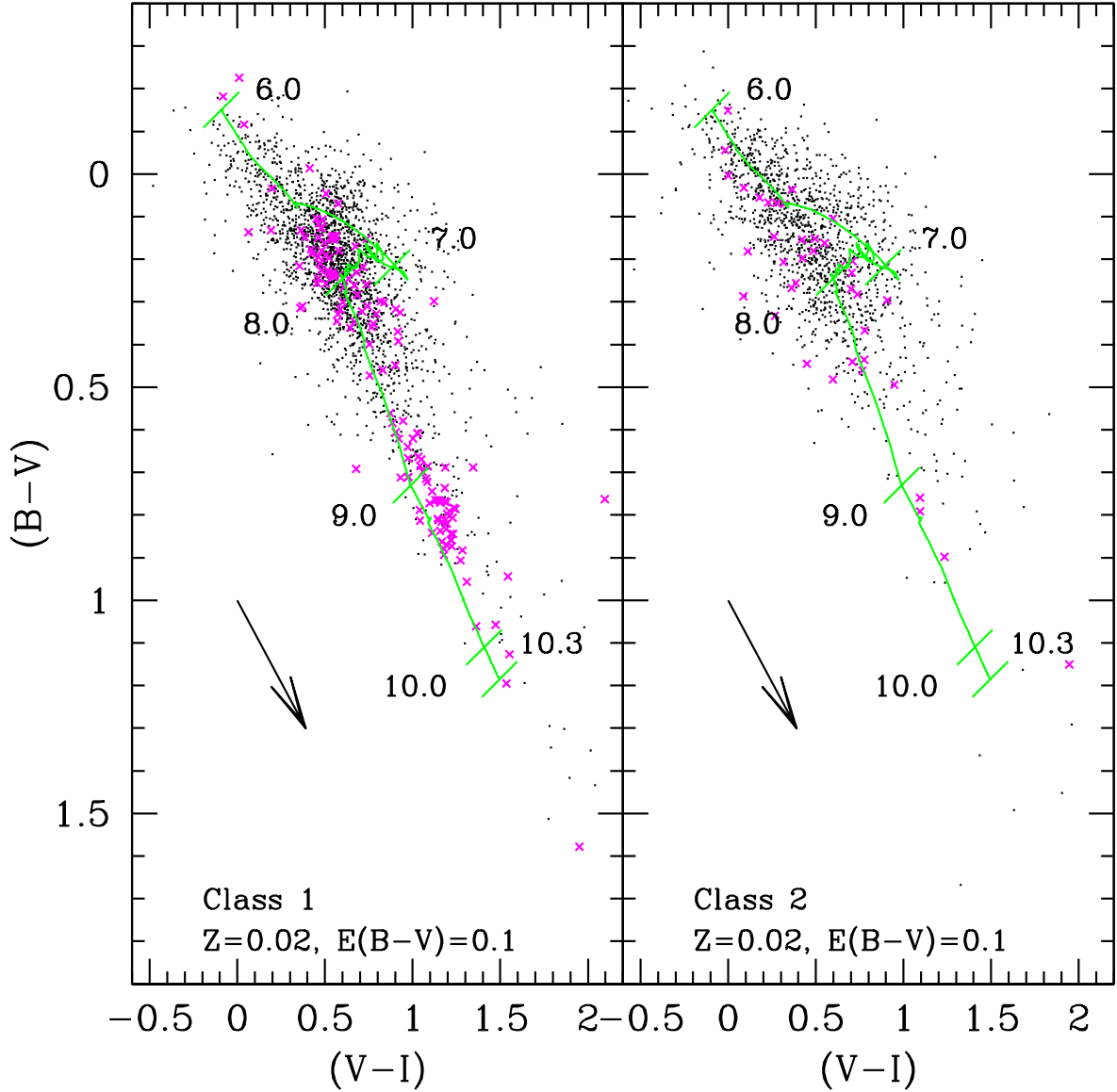} \caption{
 $(V-I)$ and $(B-V)$ color-color diagrams of Class 1 star clusters (left panel)
 and Class 2 star clusters (right panel) with $V<23$ mag in M51.
 The crosses represent the star clusters located farther than $200\arcsec$
 from the NGC 5194 center. A theoretical evolutionary track with $Z=0.02$
 by \citet{bc03} is overlaid after applying the foreground reddening of $E(B-V)=0.035$
 and the internal reddening of $E(B-V)=0.1$. Numbers that run from 6.0 to 10.3
 with tick marks on the track represent log(age).
 The arrow in each panel shows the reddening direction.
\label{ccd}}
\end{figure}

Figure \ref{ccd} shows the $(V-I)$ vs. $(B-V)$ color-color
diagrams (CCD) of Class 1 and Class 2 star clusters. Theoretical
evolutionary tracks for the simple stellar population of
\citet{bc03} with $Z = 0.02$ and $E(B-V) = 0.1$ for internal
reddening are also overlaid on these diagrams. It is evident that
most Class 1 and Class 2 star clusters are as old as $10^6 \sim
10^8$ yrs. On the other hand, red star clusters with $(B-V)>0.5$
and $(V-I)>0.8$ are considered to be older than $6 \times 10^8$
yrs and some of them might be highly reddened younger objects.
These old star clusters mostly belong to Class 1 and there are
only few old Class 2 clusters. The separation of the population of
young and old clusters is seen explicitly for star clusters
located in $d>200\arcsec$ from the NGC 5194 center (marked by
crosses in Figure \ref{ccd}). That is, young star clusters with
$\leq 10^8$ yrs are found on the blue side of $(B-V)=0.5$ and
$(V-I)=0.8$ and old star clusters with $\geq 10^9$ yrs are found
on the red side. A more detailed quantitative analysis of the star
cluster ages using the theoretical population synthesis model will
be presented in our forthcoming paper \citep{hwa07b}.

\section{Spatial Distribution of Star Clusters}
\label{spatdist}

\begin{figure}
\plotone{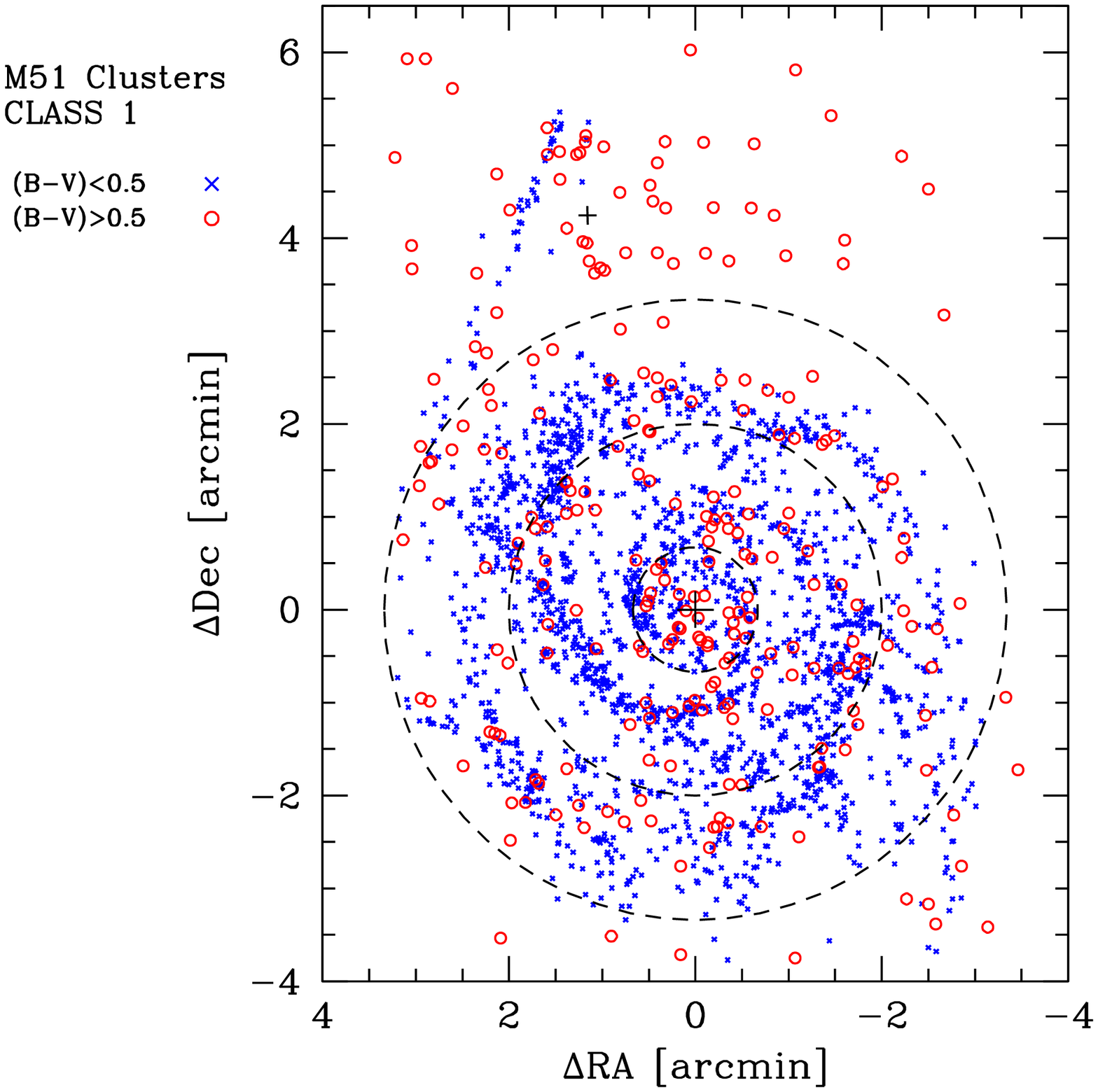} \caption{
 Spatial distribution of Class 1 star clusters in M51.
 The large circles in dashed lines indicate the concentric locations
 of $d=40\arcsec$, $120\arcsec$, and $200\arcsec$. The two large crosses
 represent the centers of NGC 5194 and NGC 5195, respectively.
 The blue clusters with $(B-V)<0.5$ are marked by small crosses,
 and the red clusters with $(B-V)>0.5$ are plotted by open
 circles. It is evident that the blue clusters are tightly
 associated with the spiral arms, while the red clusters are distributed
 more uniformly over the entire M51 field.
\label{spatc1}}
\end{figure}

\begin{figure}
\plotone{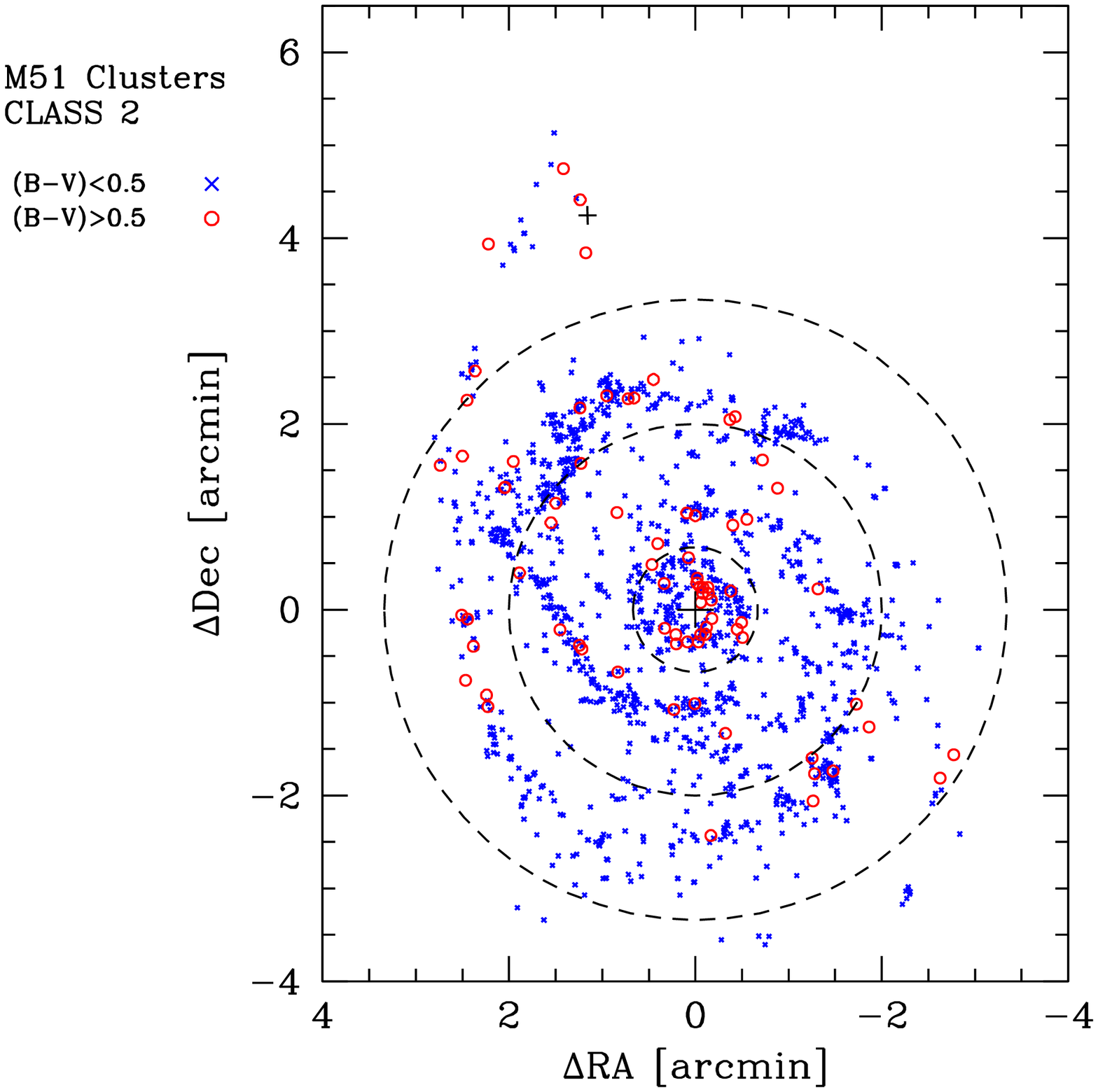} \caption{
 Spatial distribution of Class 2 star clusters in M51.
 The large circles in dashed lines indicate the concentric locations
 of $d=40\arcsec$, $120\arcsec$, and $200\arcsec$. The two large crosses
 represent the centers of NGC 5194 and NGC 5195, respectively.
 The blue clusters with $(B-V)<0.5$ are marked by small crosses,
 and the red clusters with $(B-V)>0.5$ are plotted by open
 circles. It is seen that there are very few red Class 2 clusters
 in M51.
\label{spatc2}}
\end{figure}

Figures \ref{spatc1} and \ref{spatc2} show the spatial
distribution of Class 1 and Class 2 star clusters in M51. The most
striking impression given by these figures is that star clusters
are mostly distributed in and around the two grand design spiral
arms of NGC 5194, regardless of their classification. However,
Class 2 star clusters are more closely associated with the spiral
arms than Class 1 star clusters. As a good example, the width of
the southeastern arm of M51, located at about $80 \arcsec$ from
the galaxy center, can be compared between the two classes: it is
wider for Class 1 than that for Class 2 star clusters by more than
double. It is also clear that there are more inter-arm star
clusters in Class 1 than in Class 2.

The spatial distribution of Class 1 and Class 2 star clusters
shows that the structure of the spiral arms can be traced with
star clusters as effectively as the HII regions
\citep{ran92a,lee07}, CO or HI gases \citep{rot90,schu07} in
galaxies. Therefore, star clusters can be another probe for studying the
spiral structure. The spatial distribution of Class 2 star
clusters suggests that the visual classification of star clusters
based on the morphological shape and the existence of any nearby
neighbors is efficient to select young star clusters without
utilizing any color information. However, it is necessary to use
the color information to select old globular-like star clusters.

Figure \ref{spatc1} shows that Class 1 star clusters with
different colors display different spatial distributions. The red
clusters with $(B-V)>0.5$ that are expected to be older than $6
\times 10^8$ yrs, as shown Figure \ref{ccd} are found to be
distributed rather uniformly over NGC 5194 as well as NGC 5195. On
the other hand, the blue clusters with $(B-V)<0.5$ are relatively
closely associated with the spiral arms of NGC 5194. It also shows
that the red clusters in the region of $d>200\arcsec$ are
scattered over a wide area around NGC 5195, while the blue
clusters in the same region are found only near the spiral arms
extended from NGC 5194. However, there is no evident 
difference in spatial distribution of the blue and red Class 2 clusters,
as shown in Figure \ref{spatc2}. All red Class 2 clusters
appear to be associated with the spiral arms, while only some red Class 1 clusters do.
This difference in the spatial distribution of the
red Class 1 and the red Class 2 clusters suggests that the latter may
be highly reddened objects.

\begin{figure}
 \plotone{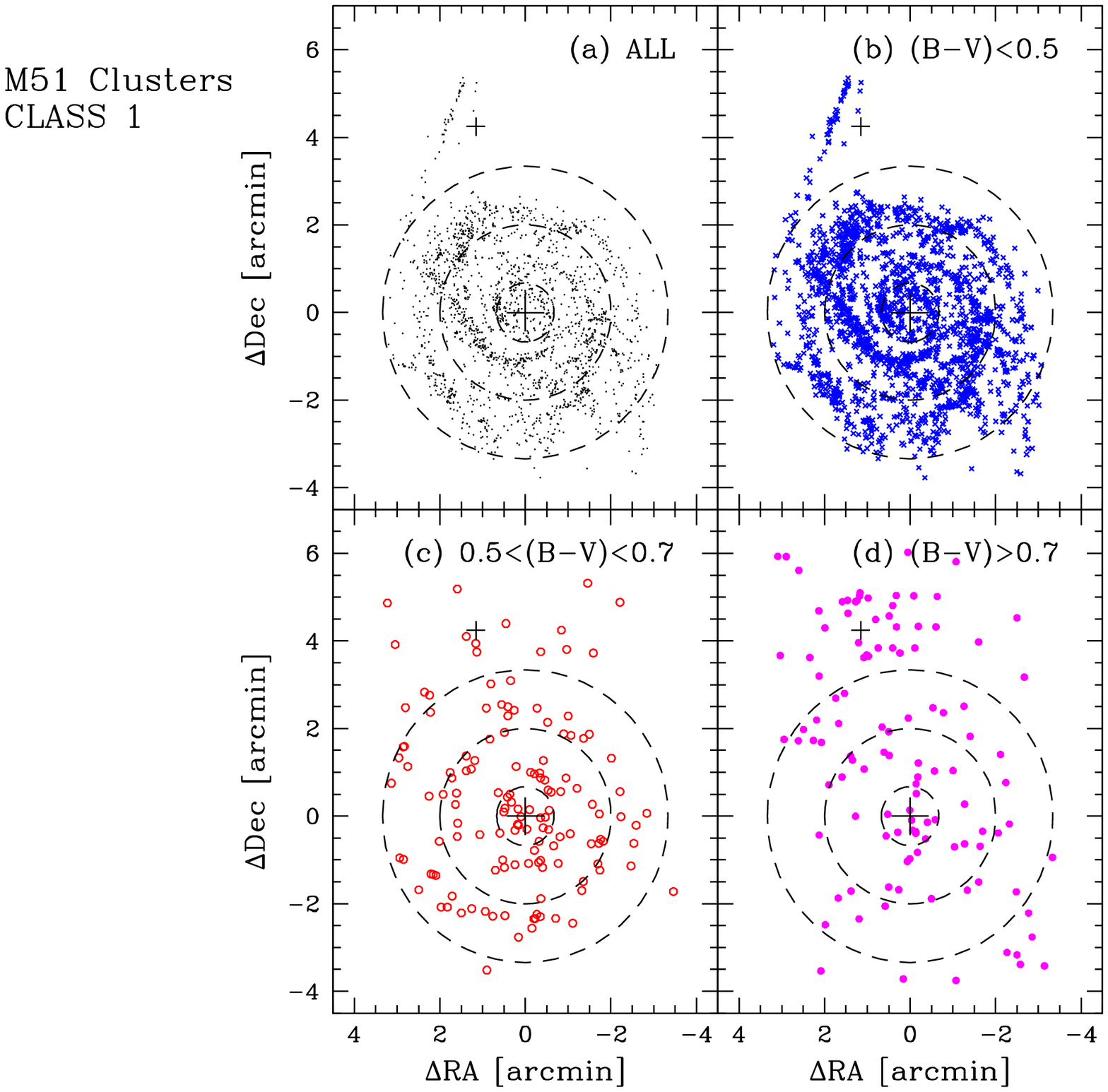}
 \caption{Spatial distribution of Class 1 star clusters with three different $(B-V)$ colors:
  star clusters with $(B-V)<0.5$ (cross) in panel (b), star clusters with $0.5<(B-V)<0.7$
  (open circle) in panel (c), and star clusters with $(B-V)>0.7$ (filled circle) in panel (d).
  Panel (a) shows the spatial distribution of all Class 1 star clusters in dots.
  In each panel, the large circles in dashed lines indicate the concentric locations
 of $d=40\arcsec$, $120\arcsec$, and $200\arcsec$, and the large crosses
 represent the centers of NGC 5194 and NGC 5195, respectively.}
 \label{spatc4}
\end{figure}

The red Class 1 clusters with $(B-V)>0.5$ include very
old star clusters with $t\geq 10^9$ yrs as well as intermediate
age clusters with $10^8 \leq t < 10^9$ yrs, as suggested in Figure
\ref{ccd}. Figure \ref{spatc4} shows the spatial distribution of
the very red Class 1 clusters with $(B-V)>0.7$ (panel (d)) as
compared to those of the other Class 1 clusters with $(B-V)<0.5$
(panel (b)) and $0.5<(B-V)<0.7$ (panel (c)). It appears that the
spatial distribution of the star clusters with $0.5<(B-V)<0.7$ is
similar to that of the blue clusters with $(B-V)<0.5$, indicating
the shape of the spiral arms. On the other hand, it is 
difficult to find any spiral-like pattern in the spatial
distribution of very red clusters with $(B-V)>0.7$. This suggests
that these very red Class 1 clusters with $(B-V)>0.7$ may belong
to the old halo of the M51 system. It is also possible that some
red Class 1 clusters may be old disk clusters of NGC 5194 and/or
NGC 5195.

\begin{figure}
 \plotone{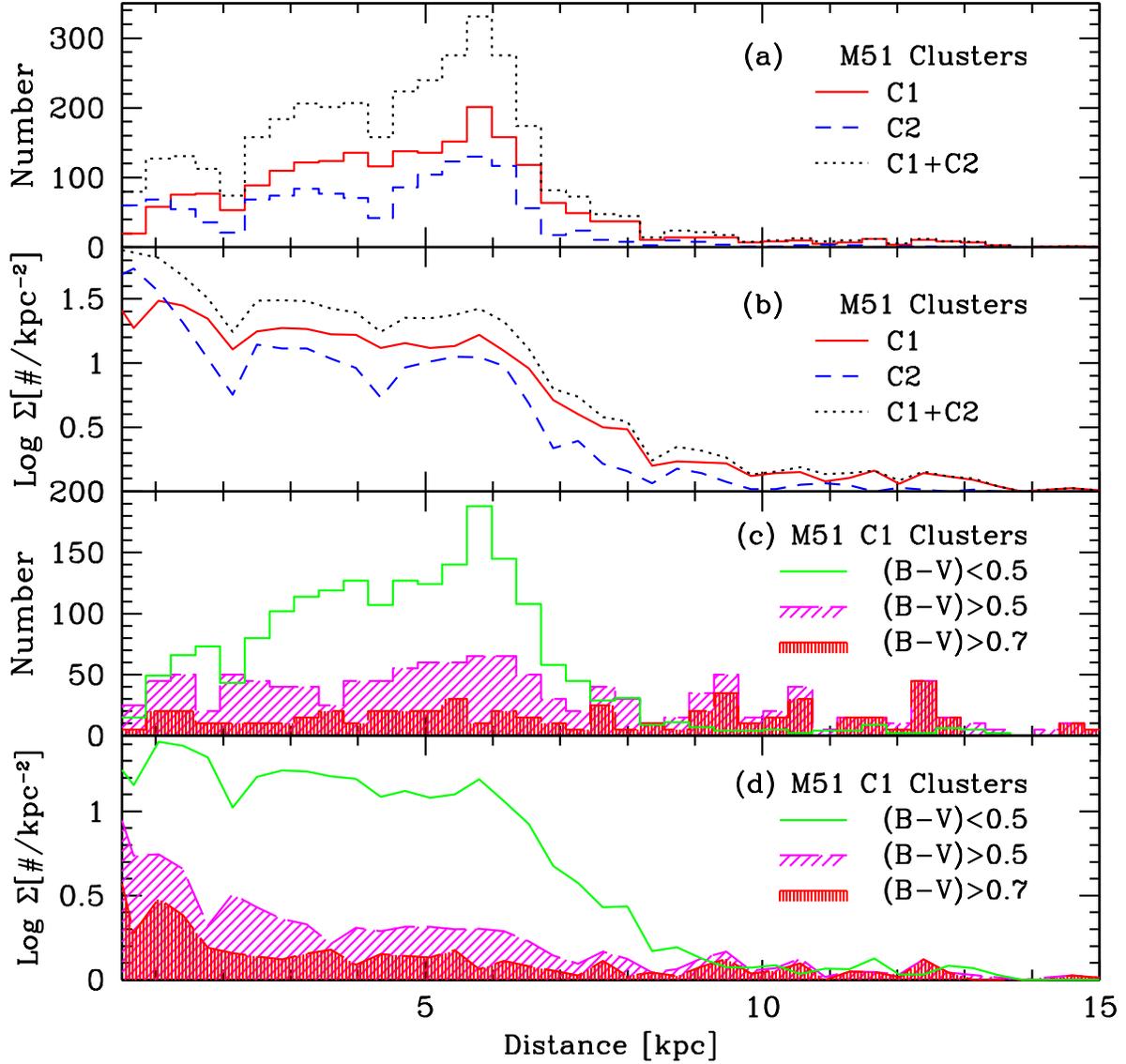}
 \caption{
  Panels (a) and (b): the radial distribution of Class 1 (solid line) and Class 2 (dashed line)
  star clusters in M51 in the number distribution (panel a) and the surface number
  density distribution (panel b). The dotted line represents the number of
  all star clusters (Classes 1 and 2).
  Panels (c) and (d): the radial distribution of the blue Class 1 clusters with $(B-V)<0.5$
  (solid line), the red Class 1 clusters with $(B-V)>0.5$ (slantly shaded), and
  the very red Class 1 clusters with $(B-V)>0.7$ (vertically shaded)
  in the number distribution (panel (c)) and the surface number density distribution
  (panel (d)).
  Please note that the histograms of the red and the very red clusters in panel (c) are
  multiplied by five for a clear presentation.
}
\label{spatc3}
\end{figure}

Figures \ref{spatc3}(a) and (b) show the radial number
distributions for Class 1 and Class 2 star clusters. For Class 2
star clusters, there are peaks centered at about 3 and
6 kpc, ($74 \arcsec$ and $147 \arcsec$), respectively. The locations of these peaks
roughly coincide with the concentric distances of the spiral arms
of NGC 5194, which suggests a strong correlation of Class 2
clusters with the spiral arms. However, for Class 1 star clusters,
the number of clusters rises rather monotonically until $D \approx
6$ kpc. The rapid decrease in the number of clusters in $D > 6$
kpc is shared by Class 1 and Class 2 clusters, since this is where
the major spiral arms meet their outer boundary. The surface
number density of the clusters, shown in panel (b) of Figure
\ref{spatc3}, appears to display three distinct parts: a central
peak ($D < 2$ kpc), a mid-plateau ($2 < D < 6$ kpc), and an abrupt
decline ($D > 6$ kpc). This is in good agreement with the result
of \citet{sch06}.

Figures \ref{spatc3}(c) and (d) show the radial
distributions of three Class 1 cluster populations with different
colors. It is clear that the blue Class 1 clusters with
$(B-V)<0.5$ are centrally concentrated within $D = 7$ kpc from the
NGC 5194 center, and the surface number density drops sharply at
$D \approx 6$ kpc. On the other hand, the red Class 1 clusters
with $(B-V)>0.5$ show a rather constant number distribution up to
$D \approx 6$ kpc, which leads to a smoothly declining surface
number density profile. The very red Class 1 clusters with
$(B-V)>0.7$, however, display a flat number distribution across
the entire survey field of M51. The surface number density profile
of these red clusters shows a small enhancement only at the
center, and a flat distribution in the other regions. This may be
caused by the preferential disruption of clusters inside the disk.
These differences in the spatial and radial distribution of the
star clusters, as shown in Figures \ref{spatc1} $\sim$
\ref{spatc3}, suggest that the red Class 1 star clusters are
composed of old halo population as well as old disk population,
survivors of the cluster disruption, and that blue star clusters
(both Class 1 and 2) are young disk populations.

\section{Size Distribution of Star Clusters}

We estimated the sizes of star clusters in M51
using ISHAPE and adopted the size measurements with $(\chi^2/\chi^2_0)<0.9$.
The ratio $(\chi^2/\chi^2_0)$ indicates how
significantly the shape of a source deviates from that of the
input PSF. In most cases, however, it suggests the relative
extendedness of a source compared to the PSF. The measurements
with $(\chi^2/\chi^2_0) \geq 0.9$ mostly returned the radii of
$R_{\rm eff}<2$ pc or $1$ pixel, which intrinsically complicates the
accurate estimation of size. In particular, in case that the
empirically derived PSF is used for the size measurement, as in
this study, the implementation of this ratio helps to select only
the clusters that return reliable size measurements. However, the
adoption of this parameter tends to remove star
clusters that have as small radii as $R_{\rm eff}<2$ pc.

\begin{figure}
 \plotone{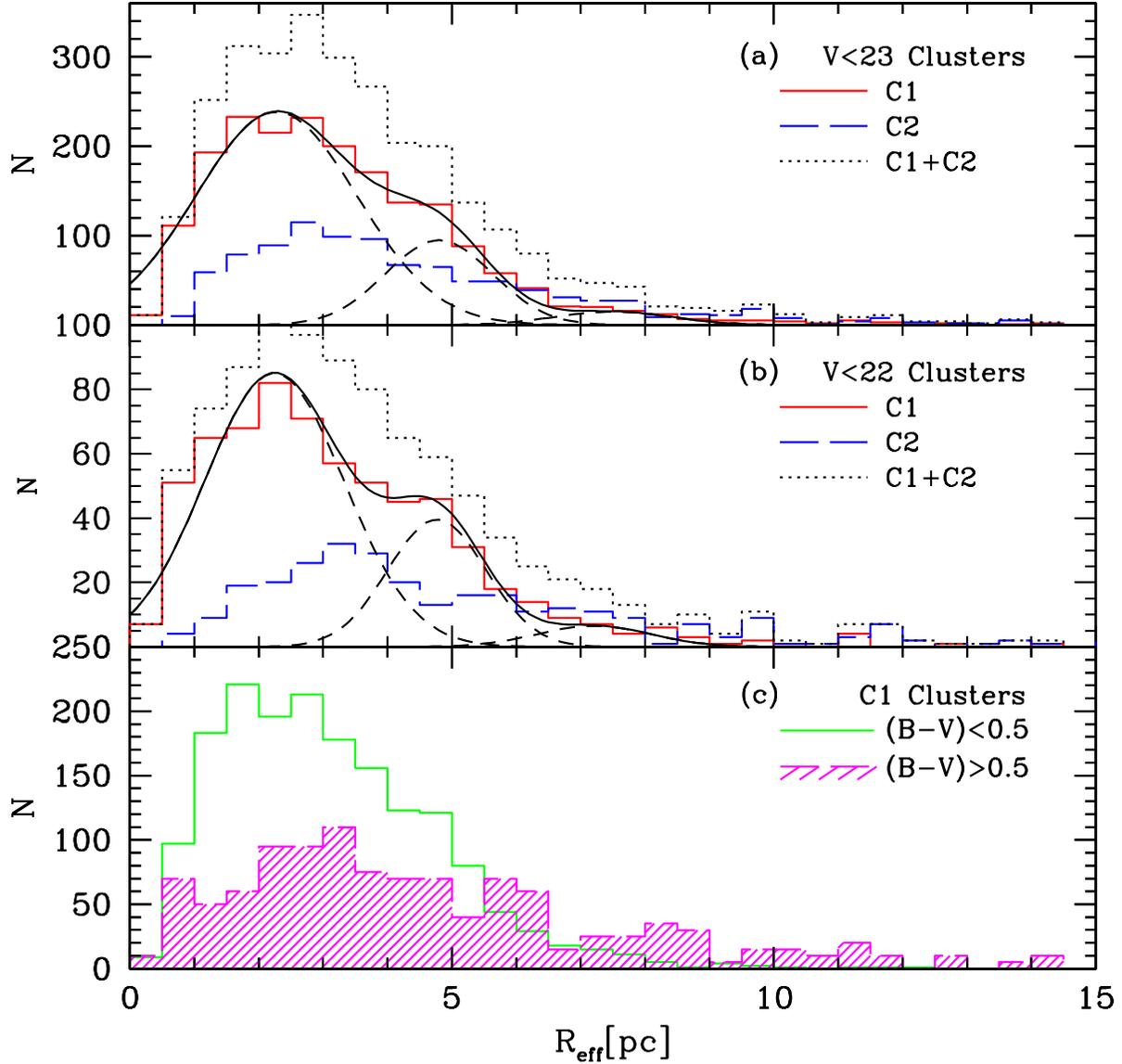}
 \caption{Panel (a) shows the effective radius $R_{eff}$ distribution
 of Class 1 (solid line) and Class 2 (long dashed line) star clusters
 with $V<23$ mag and $(\chi^2/\chi^2_0)<0.9$ in M51. The dotted line represents
 the distribution of all star clusters (Classes 1 and 2). Panel
 (b) shows the same distribution for the star clusters with $V<22$ mag.
 Gaussian profiles for Class 1 cluster size distribution are plotted in dashed curves
 and the sum of three Gaussians is shown in solid curves.
 Panel (c) shows the size distribution plot for the blue Class 1 clusters
 with $(B-V)<0.5$ (solid line) and the red Class 1 clusters with
 $(B-V)>0.5$ (shaded long dashed line).
 Please note that the histogram of the red Class 1 clusters is
 multiplied by five for a clear presentation.
}
\label{hldist}
\end{figure}

Figure \ref{hldist} shows the distribution of the effective radii
$R_{\rm eff}$ of star clusters. 
The size distributions of Class 1 clusters with $V<23$ (and $V<22$) appear to be composed of three
seperate components: a dominant peak at about 3 pc, a weaker one at about 5 pc, 
and a long much weaker tail in the larger range.
By fitting three Gaussian profiles to the data, we derive
the largest component centered at $2.27 \pm 0.03$ pc with $\sigma = 1.90$ pc, 
the second component at $4.80 \pm 0.04$ pc with $\sigma = 1.26$ pc, 
and the third one at $7.51 \pm 0.30$ pc with $\sigma = 1.38$ pc, 
as shown in Figures \ref{hldist}(a) and (b).
The first is very similar to the median size of globular clusters in the Milky Way,
which is approximately 3 pc \citep{har96}.
The second one represents the intermediate-sized star clusters and
they are mostly bluer than $(B-V)=0.5$, as shown in Figure \ref{hldist}(c).
The third one corresponds to the large star clusters described in Section \ref{lsc}
and it also includes the faint fuzzy clusters \citep{lee05, hwa06}.
It is also worth noting that this size distribution of clusters in M51 can be biased against very
compact clusters due to the cluster selection criteria and the size measurement constraints adopted. 

Figure \ref{hldist}(c)
shows that the blue Class 1 clusters with $(B-V)<0.5$
are mostly smaller than $R_{\rm eff}=5$ pc, and their size
distribution is consistent with that of the entire cluster sample.
However, the red Class 1 clusters with $(B-V)>0.5$ display a large
range of radius from $R_{\rm eff} \approx 1$ to $10$ pc or
larger and their size distribution may be represented with a
very broad single Gaussian centered at $3.66 \pm 0.30$ pc with $\sigma = 3.53$ pc. 
One more point to note is that the slope of the size
distribution at $R_{\rm eff}>3$ pc region is quite shallower for Class
2 than that for Class 1 clusters. It is the same case with the red
Class 1 clusters that display a size distribution slope shallower
than that of the blue Class 1 clusters. This suggests that there
are a larger number of large Class 2 and red Class 1 clusters.

\begin{figure}
 \plotone{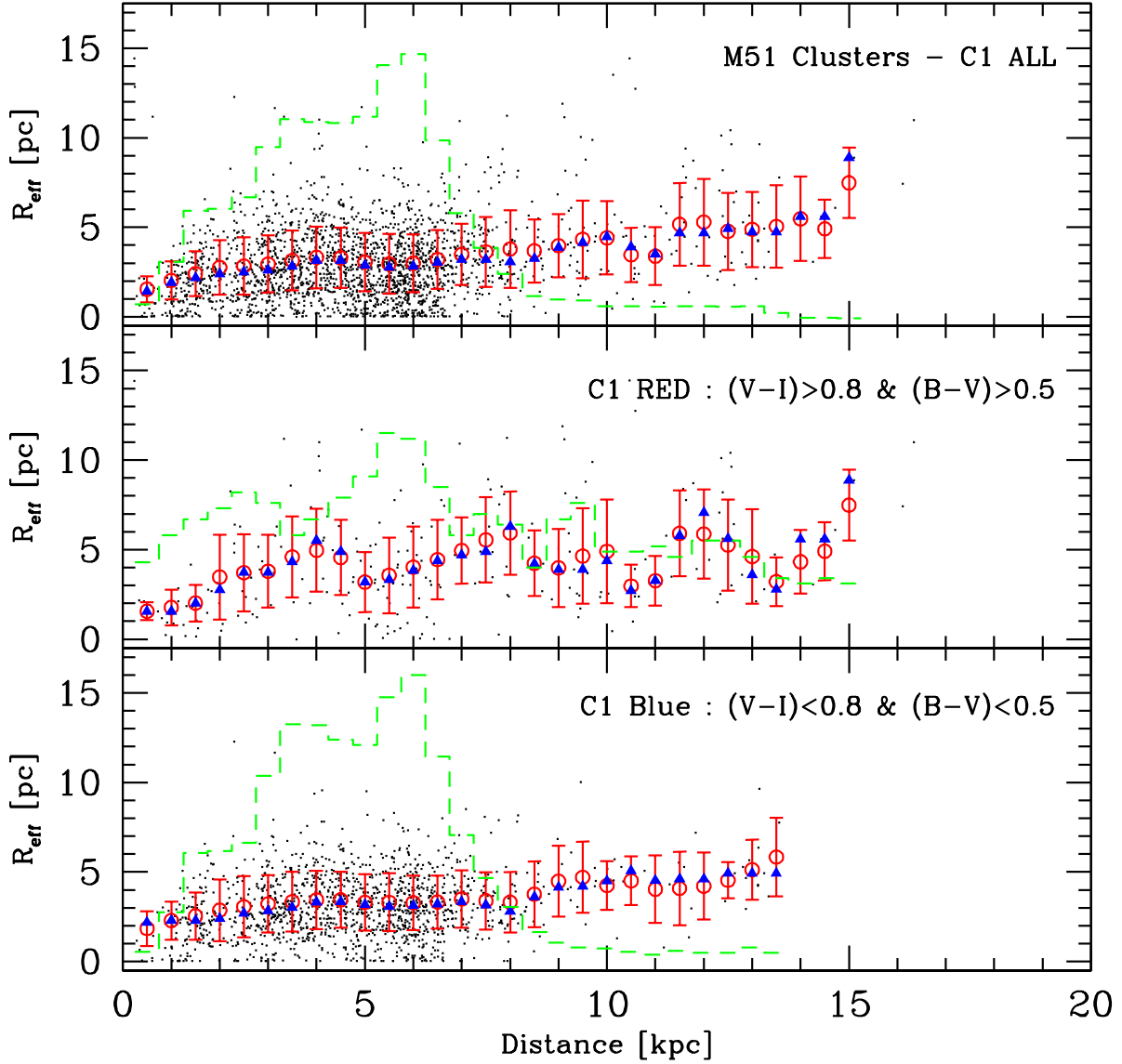}
 \caption{Size variation of the  
Class 1 clusters along the galactocentric distance of NGC 5194. The upper panel shows the
distribution of all Class 1 clusters and the mid panel shows the
same distribution of the red Class 1 clusters with $(V-I)>0.8$ and
$(B-V)>0.5$, while the lower panel displays that of the blue Class
1 clusters with $(V-I)<0.8$ and $(B-V)<0.5$. In each panel, a
circle represents the mean effective radius of clusters in each
bin, while a triangle represents the median. The histogram in
dashed lines shows the number distribution of Class 1 star clusters along the
galactocentric distance. } \label{sizedist}
\end{figure}

For globular clusters in the Milky Way and LMC, it is known that
there is a correlation between the upper limit of the effective
radii ($R_{\rm eff}$) and the galactocentric distance ($D$)
\citep{vdb04}. We have investigated any such correlation in M51
using Class 1 clusters. 
Figure \ref{sizedist} shows the distribution of $R_{\rm eff}$ 
vs. $D$ of Class 1 star clusters. 
It shows that the effective radii of clusters increase as the
galactocentric distance increases, as also suggested by
\citet{sch07}. The median $R_{\rm eff}$ of the star clusters located
in $D<7$ kpc is about $2.78 \pm 1.64$ pc, but the median $R_{\rm eff}$
in $D>8$ kpc is about $4.32 \pm 2.09$ pc. Although the differences
in the radii are smaller than the statistical errors, the size
distribution of the star clusters in $D < 7$ kpc and in $D > 8$
kpc, as shown in the upper panel of Figure \ref{sizedist}, is
clearly different. This is in part due to the lack of small
clusters in the outer region since the number of star clusters
with $R_{\rm eff} < 5$ pc is about 1,500 for $D < 7$ kpc and about 100
for $D > 8$ kpc whereas the number of star clusters with $R_{\rm eff}
> 7$ pc is about 50 for $D < 7$ kpc and about 30 for $D > 8$ kpc.
Moreover, the number of very large star clusters with $10< R_{\rm eff}
<20$ pc is about 10 in both the inner and outer regions.
Therefore, in the inner region of the spiral arms, there are a large number
of typical small star clusters, while large star clusters are
evenly distributed over the projected plane of M51.

Figure \ref{sizedist}(c) shows the size
distribution of the red clusters with $(B-V)>0.5$ and $(V-I)>0.8$ and 
and the blue clusters with $(B-V)<0.5$ and $(V-I)<0.8$. These
color criteria are consistent with those used to separate Galactic
globular clusters. 
The size distribution of the red clusters is much broader than that
of the blue clusters. 
The median $R_{\rm eff}$ of the red clusters in $D<7$
kpc is about $3.45 \pm 2.16$ pc, while that of the blue clusters
is about $3.04 \pm 1.59$ pc. Another noteworthy point is that
there are many more red clusters with $R_{\rm eff} > 10$ pc than blue
clusters. The number of the large red clusters is 16, while that
of the large blue clusters is only three.

\section{Large Star Clusters}
\label{lsc}

Apart from typical star clusters with $R_{\rm eff} \approx 3$ pc,
large star clusters with $R_{\rm eff} > 7$ pc are being discovered
recently: e.g., faint fuzzies in NGC 1023 and 3384
\citep{larb00,bro02}, extended star clusters in M31
\citep{hux05,mac06}, remote and extended star clusters in NGC 6822
\citep{hwa05}, diffuse star clusters in the Virgo cluster
\citep{pen06}, and so on. These large star clusters are being
considered as a new family of star cluster population. In M51, it
has been already reported by \citet{cha04} that there are more
large star clusters compared to other nearby late-type galaxies.
\citet{lee05} later discovered that there are faint fuzzy clusters
around NGC 5195, a companion galaxy of NGC 5194. Based on $HST$
ACS data, \citet{hwa06} identified 49 faint fuzzy clusters around
NGC 5195 and also reported that these large clusters display a
different spatial distribution pattern, compared to the compact
clusters with similar colors, implying a different origin for
these faint fuzzy clusters.

\begin{figure}
 \plotone{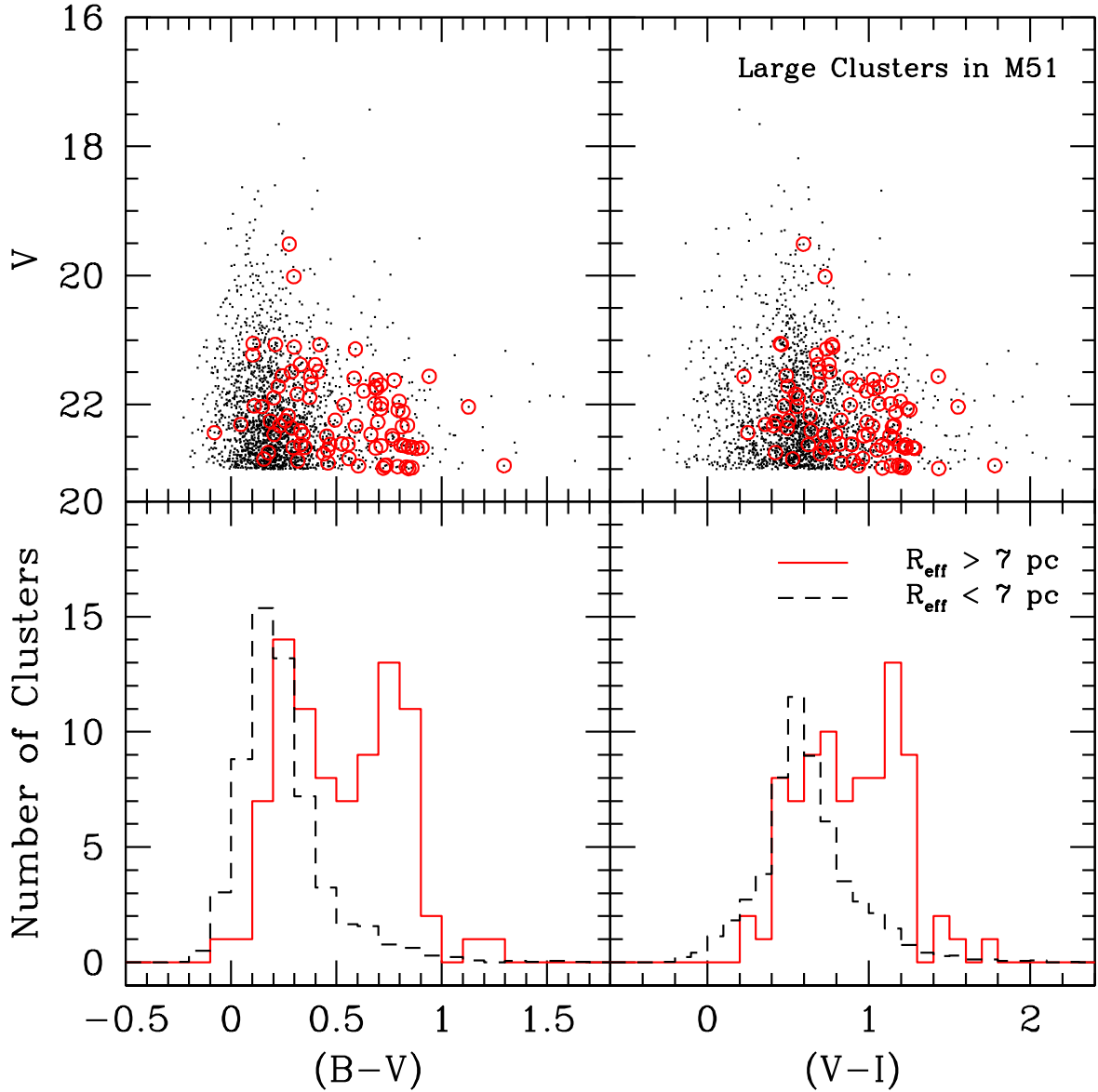}
 \caption{Upper panels: $(B-V)$ vs. $V$ and $(V-I)$ vs. $V$ color-magnitude diagrams (CMD)
of Class 1 clusters (dots) and LSCs (circles) in M51. Lower
panels: $(B-V)$ and $(V-I)$ color distribution of Class 1 clusters
with $R_{\rm eff} < 7$ pc (dashed lines) and LSCs with $R_{\rm eff} > 7$
pc (solid lines) in M51. Please note that the histograms in dashed
lines display $3\%$ of the real values for a better comparison.
See the text for details.} \label{lgccmd}
\end{figure}

We define large star clusters (hereafter LSC) as star clusters
with $R_{\rm eff}>7$ pc without any reference to the luminosities nor
to the colors. The faint fuzzy clusters found in NGC 5195
\citep{hwa06} belong to this LSC. The photometric properties of
the LSCs are displayed in Figure \ref{lgccmd}. The color-magnitude
diagrams in the upper panel of Figure \ref{lgccmd} show that the
LSCs are mostly fainter than $V = 21$ mag and only two LSCs are
brighter than this. Image inspection revealed that these two
bright LSCs are very young clusters located in actively
star-forming regions with profuse $H\alpha$ emissions, possibly
constituting the so-called star cluster complexes recently
reported by \citet{bas05b}. The lower panels of Figure
\ref{lgccmd} show the $(B-V)$ and $(V-I)$ color histograms of LSCs
in solid lines and Class 1 clusters with $R_{\rm eff}<7$ pc in dashed
lines. Please note that the histograms for Class 1 clusters with
$R_{\rm eff}<7$ pc are scaled down to 3\% of their real values for a
better comparison with those of the LSCs. It is clear that the
color distributions for LSCs and Class 1 clusters with $R_{\rm eff}<7$
pc exhibit different features: the former displays two distinct
peaks, while the latter shows only one blue peak. Another
noteworthy point is that there are very few LSCs with $(B-V)<0.1
\sim 0.2$ as compared to Class 1 clusters with $R_{\rm eff}<7$ pc.

\begin{figure}
 \plotone{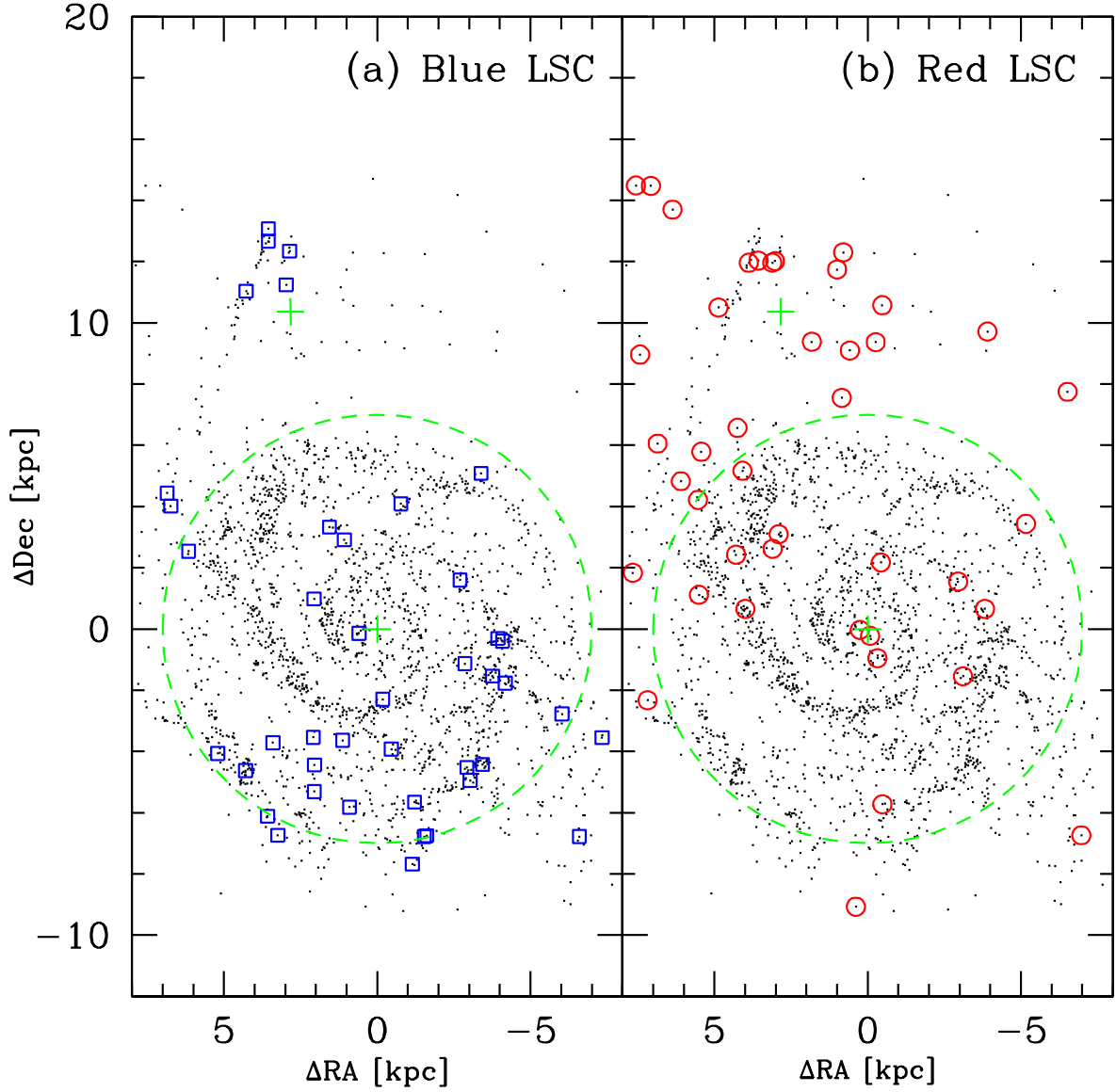}
 \caption{Spatial distribution of LSCs in M51.
The blue LSCs with $(B-V)<0.5$ are marked by squares in panel (a)
and the red LSCs with $(B-V)>0.5$ are by circles in panel (b),
while Class 1 clusters are plotted in dots in each panel. The
crosses in each panel indicate the centers of NGC 5194 and NGC
5195, respectively. A large circle with $D=7$ kpc ($d \simeq 172\arcsec$)
is drawn to display the approximate extent of the
disk of NGC 5194. } \label{lgcspat}
\end{figure}

The $(B-V)$ and $(V-I)$ color distributions of the LSCs shown in
the lower panel of Figure \ref{lgccmd} suggest that there are two
distinct populations in the LSCs: blue population and red
population separated by the colors of $(B-V)=0.5$ and $(V-I)=0.8$.
The result of the image inspection suggests that most of the blue
LSCs appear to be young clusters associated with other blue
clusters. The spatial distribution shown in Figure \ref{lgcspat}(a)
also indicates that about half of these blue LSCs
are found near the spiral arms of NGC 5194.
However, these blue LSCs tend to be found more in the
southern region of NGC 5194 disk than in the northern disk.
On the other hand, the red LSCs marked by circles in panel (b) of
Figure \ref{lgcspat} are found to be more concentrated in the
northeastern arm and near the NGC 5194 nucleus than those in the
southern disk. Moreover, these red LSCs are widely distributed up
to the field of NGC 5195. It is possible that some of these
red LSCs in the NGC 5194 disk may be reddened blue LSCs.

The red LSCs include a significant number of faint fuzzy clusters
that satisfy the selection criteria defined in \citet{hwa06}.
Among the 42 red LSCs, there are 27 faint fuzzy clusters,
including 15 previously identified in \citet{hwa06}. These
clusters are noted by `FF' comments in the final Class 1 cluster
catalog presented in Section \ref{cat}. Figure \ref{ffspat} shows
the spatial distribution of these faint fuzzy clusters (panel (b))
in comparison with compact red clusters (panel (a)) that are
smaller than $R_{\rm eff} = 7$ pc but have the same $(B-V)$ and
$(V-I)$ colors as the faint fuzzies. It is clear that the faint
fuzzy clusters are more concentrated around NGC 5195 than around NGC
5194. Even the faint fuzzies around NGC 5194 appear to be
distributed along the northeast-southwest direction across the
disk. Interestingly, this direction is the same as that found for
the faint fuzzy clusters associated with NGC 5195 \citep{hwa06}.
However, the compact red clusters are distributed rather uniformly
over the entire M51 field, which indicates that the compact red
clusters and faint fuzzies are different populations.

\section{Discussion}
\label{discuss}

We have detected and selected 2,234 Class 1 and 1,388 Class 2 star
clusters in M51 through the visual inspection of about 8,400
sources with $V<23$ mag. The final sample of Class 1 star clusters
that have a circular shape and no prominent neighbor is compiled
into a catalog. This star cluster catalog of M51 has been
made using a homogeneous dataset and a consistent selection
method over the entire field of M51 including both NGC 5194 and
5195. Unlike the star cluster catalog generated using the
automated method, this catalog does not provide a complete list of
star clusters in M51. This catalog may suffer from the
incompleteness and the subjectivity that are unavoidable in a
visual classification process. However, this catalog does provide
a least contaminated list of star clusters in M51, which is
essential for various studies on star clusters. We have used this
cluster catalog and have found some interesting photometric and
physical properties of star clusters in M51. This star cluster
catalog will be used for a future study on the age distribution of
star clusters in M51 \citep{hwa07b}.

We have found that the star clusters in M51 are mostly located
near the spiral arms of NGC 5194 regardless of their morphologies.
However, Class 2 star clusters appear to be more closely
associated with the spiral arms than Class 1 star clusters, some
of which are also found between the spiral arms. Another important
fact is that a significant number of Class 1 star clusters and
nearly all of Class 2 star clusters are bluer than $(B-V)=0.5$ and
$(V-I)=0.8$, which suggests that most of the star clusters are
young clusters. However, in the outer space of the M51 system,
there are more red star clusters with $(B-V)>0.5$ and $(V-I)>0.8$,
which is in contrast to the inner disk region of NGC 5194. The
$(V-I)$ and $(B-V)$ color-color diagrams show that these red
clusters are expected to be old ($t \geq 6 \times 10^8$ yrs). This
suggests that even though most star clusters in M51 are blue and
young clusters that are formed less than $t \approx 10^8$ yrs ago,
there are also some red and old star clusters that should have
been formed no less than $t \approx 10^9$ yrs ago, that is, before
the epoch of the dynamical interactions between NGC 5194 and 5195.

\begin{figure}
 \plotone{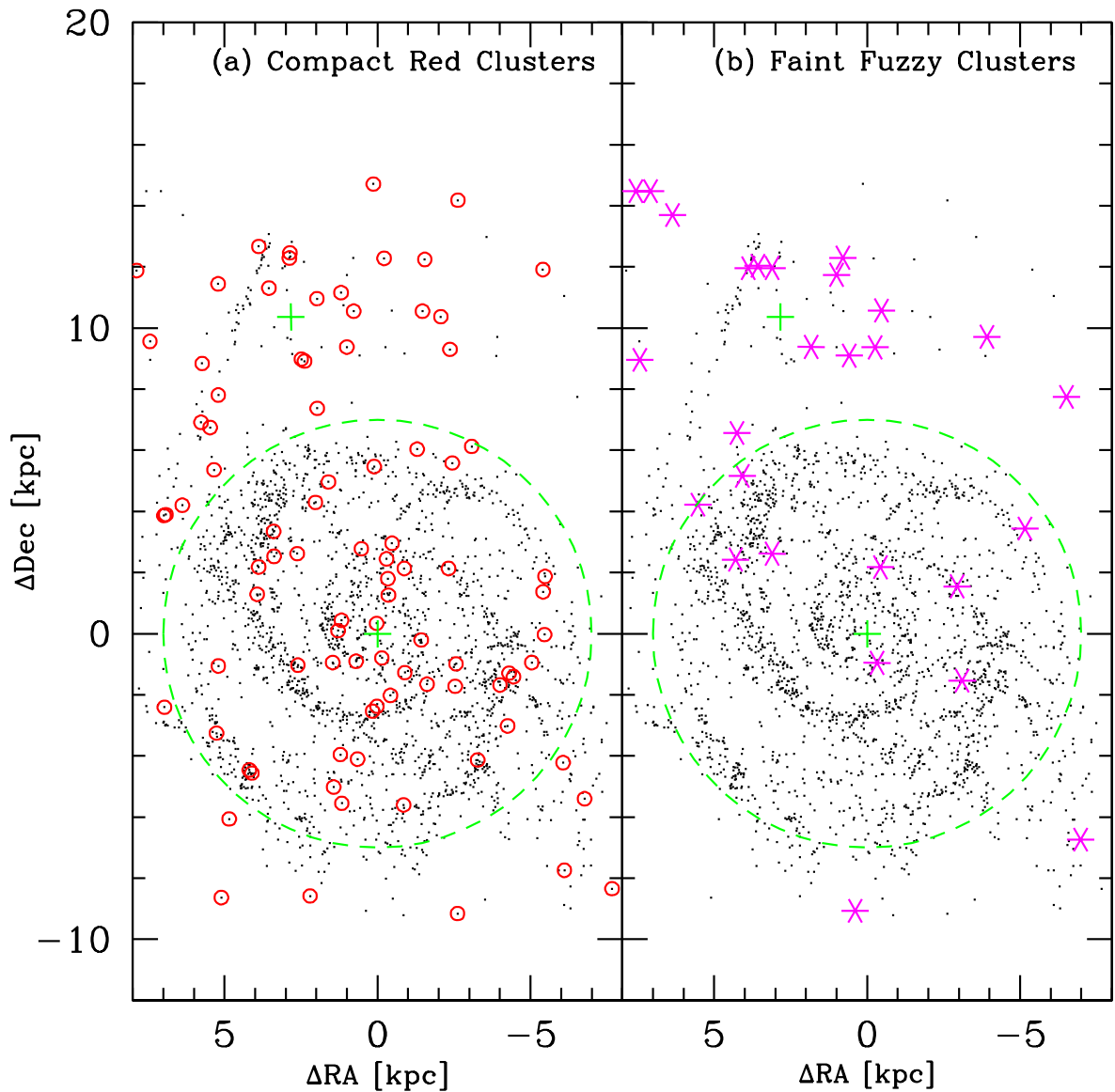}
 \caption{Spatial distribution of the compact red clusters (panel (a))
and the faint fuzzy clusters (panel (b)) in M51 in comparison with that
of Class 1 clusters plotted in dots. The crosses in each panel
indicate the centers of NGC 5194 and NGC 5195, respectively. A
circle with $D=7$ kpc ($d \simeq 172\arcsec$) is drawn to
display the approximate extent of the disk of NGC 5194.}
\label{ffspat}
\end{figure}

Old star clusters in M51 are relatively understudied in the
previous studies as compared to young and blue clusters. In a
recent study on the globular cluster systems of nearby late-type galaxies, the
globular cluster specific frequency $S_N$ was calculated by \citet{cha04}. They
detected 34 globular clusters in M51 using the $HST$ WFPC2 archive data and
calculated the globular cluster specific frequency to be $S_N = 0.5 \pm 0.1$. We
assume that the compact red clusters defined in Section \ref{lsc}
and shown in Figure \ref{ffspat} are globular clusters, and we
calculate the globular cluster specific frequency of M51 using the same method
adopted by \citet{cha04}. The number of the compact red clusters
with $M_V <-7.2$ mag is 48 and the total number of these clusters
would be 96 if we assume that the luminosity function has a
symmetric lognormal shape. The corresponding globular cluster specific frequency
is $S_N = N_{GC} \times 10^{+0.4(M_V + 15)} \approx 0.20$ if we
adopt $M_V = -21.68$ for M51, as used by \citet{cha04}. However,
the compact red clusters are a subset of real globular clusters
because they are selected based on the same color criteria used
for the selection of faint fuzzy clusters, that is, $0.6 < (B-V) <
1.1$ and $1.0 < (V-I) < 1.5$. Therefore, if we relax these
criteria and adopt more widely accepted colors for globular
cluster selection, that is, $(B-V)>0.5$ and $(V-I)>0.8$, then the
number of the selected clusters with $M_V<-7.2$ mag increases to
91, which leads to $S_N \approx 0.39$. This is a rather small
value compared to the result of \citet{cha04}. However, we do not
apply any type of incompleteness correction to our data. One more
thing to consider when we compare these two results is that
globular clusters selected by \citet{cha04} include large clusters
with $R_{\rm eff}>7$ pc while our definition of globular clusters only
allows clusters with $R_{\rm eff}<7$ pc. Including the large clusters
into the globular cluster sample would increase the globular cluster specific
frequency by about 0.10, yielding $S_N \approx 0.48$, which is in
good agreement with the estimation by \citet{cha04}.

The size distribution of star clusters in M51 shows that most of
the star clusters selected in this study are as large as $R_{\rm eff}
\approx 2.0 \sim 3.3$ pc with a median $R_{\rm eff} = 2.92$ pc. This
value is very similar to that of Galactic globular clusters, which
is $R_{\rm eff} \sim 3.3$ pc \citep{har96}. However, this size
distribution of clusters in M51 can be biased against very compact
clusters due to the cluster selection criteria and the size
measurement constraints adopted in this study. Recently,
\citet{sch07} reported relatively small radii with a median
$R_{\rm eff} = 2.1$ pc for their cluster candidates in M51 based on
the same Hubble Heritage Data. The cluster size for one extremely
large cluster ID212995 noted in their paper (ID54775 in our
catalog) also shows a small difference: 21.6 pc in \citet{sch07}
and 22.79 pc in this study. The difference in the star cluster
size and its distribution between in this study and in
\citet{sch07} may be caused by the different cluster selection
methods and the different PSFs used in the size measurement.
Nonetheless, other characteristics displayed in the size
distribution of star clusters still appear to be common in both
studies. The logarithmic size distribution of the star clusters
between $R_{\rm eff} =3$ and $5$ pc is found to be well represented by
the slope of $-1.2$ both in this study and in Figure 14 of
\citet{sch07}. As another example, we investigated the size of
clusters with $(B-V)<0.1$ and $(B-V)>0.1$ 
following the definition by \citet{sch07}. We found that the
median $R_{\rm eff}$ for the blue clusters is 2.00 pc and that of the
red clusters is 3.24 pc, indicating that the red clusters are
systematically larger than the blue clusters. This result is
consistent with that of \citet{sch07} which shows that the median
$R_{\rm eff}$ is $1.8$ pc for the blue clusters, while it is $2.5$ pc
for the red clusters.

The investigation using the star cluster size and the
galactocentric distance has returned no obvious correlation
between the two parameters, unlike that for Galactic globular
clusters. However, the star clusters on the disk of NGC 5194
appear to be statistically smaller than the clusters off the disk.
This is qualitatively consistent with Figure 19 of \citet{sch07}.
Although it is very uncertain due to the large statistical error,
it appears that the fraction of large red clusters with
$(B-V)>0.5$ and $(V-I)>0.8$ is higher than that of large blue
clusters as shown in Figures \ref{hldist} and \ref{sizedist}. For
clusters with $R_{\rm eff}>10$ pc, there are 16 large red clusters,
while only three such blue clusters are found, which is in
striking contrast.

We have found that there are many LSCs
in M51. LSCs are mostly fainter than $V = 21$ mag and, unlike
typical Class 1 clusters with $R_{\rm eff}<7$ pc, they have two
populations with different colors: (1) the blue population with
$(B-V)<0.5$ and $(V-I)<0.8$ and (2) the red population with
$(B-V)>0.5$ and $(V-I)>0.8$. The blue and red LSCs are found to
display slightly different spatial distributions. Many red LSCs
are distributed not only over the disk of NGC 5194 but also around
the body of NGC 5195, whereas the blue LSCs are mostly located
around the NGC 5194 disk. Another interesting point is that even
the LSCs around NGC 5194 display different spatial distributions
with different colors: the red LSCs are relatively more
concentrated in the northern disk, while the blue LSCs are
concentrated in the southern disk. It is not yet clear what causes
this difference.

Some of these red LSCs are faint fuzzy clusters that satisfy the
color criteria defined in \citet{hwa06}. These faint fuzzy
clusters are preferentially found around NGC 5195, but there are
still some faint fuzzies that appear to be associated with NGC
5194. Interestingly, the spatial distribution of these faint fuzzy
clusters is very different from that of the compact red clusters,
as shown in Figure \ref{ffspat}. The compact red clusters display
rather uniform distribution over the entire field of M51. On the
other hand, the faint fuzzy clusters are distributed
preferentially along the northeast-southwest direction. In
\citet{hwa06}, the faint fuzzies in NGC 5195 are shown to display
a very similar spatial distribution that is elongated along the
same northeast-southwest direction. It is pointed out by
\citet{hwa06} that the elongated distribution of the faint fuzzies
may be a part of the debris of tidal interactions between NGC 5194
and NGC 5195. If this is the case, the spatial distribution of the
faint fuzzies over the disk of NGC 5194 could be a part of the
same debris. It could provide us with an important clue that might
reveal the passage route of a companion galaxy in the M51
system.

\section{Summary}
\label{sum}

Using the Hubble Heritage Data taken with $HST$ ACS, we 
selected about 3,600 star clusters in M51 based on their
morphological information after visually inspecting about 8,400
sources with $V<23$ mag. We present a final catalog that includes
2,224 Class 1 star clusters with a circular shape and no prominent
adjacent neighbors. 
The selected clusters are mostly found around
the spiral arms of NGC 5194. Most of the star clusters are bluer
than $(B-V)=0.5$ and $(V-I)=0.8$, suggesting that they are young
clusters with $t \leq 10^8$ yrs. However, there are red clusters
around both NGC 5194 and NGC 5195. Some of these red clusters with
$(B-V)>0.7$ and $t \geq 10^9$ yrs are considered to be a part of
the halo population. 
We derived the luminosity function of the clusters, finding that it is
fit well by a single power law 
with $\alpha = -2.59 \pm 0.03$ for the range $-10.0 <M_V < -8.0$ mag.
We found that the size distribution of the star clusters can be fit with
three Gaussian components:
a dominant component with a peak at $R_{\rm eff}=2.27 \pm 0.03$ pc with $\sigma = 1.90$ pc, 
a second component with a peak at $4.80 \pm 0.04$ pc with $\sigma = 1.26$ pc, 
and a much weaker broad one with a peak at $7.51 \pm 0.30$ pc with $\sigma = 1.38$ pc.
The first is very similar to the median size of globular clusters in the Milky Way,
The second one represents the intermediate-sized star clusters and
they are mostly bluer than $(B-V)=0.5$, 
and the third one corresponds to the large star clusters.
Among the large clusters, we found 27 faint fuzzy clusters that are distributed
over the disk of NGC 5194 as well as around the body of NGC 5195.
Interestingly, these faint fuzzy clusters are found to show an
elongated spatial distribution along the northeast-southwest
direction over the NGC 5194 disk. This spatial distribution is
very different from that of the compact red clusters that are
distributed rather uniformly over the entire field of M51. Further
studies on these clusters are needed to understand what induces
this difference as well as the origin of these different star
cluster populations.

\acknowledgements

The authors are grateful to the anonymous referee for the critical
comments and useful suggestions that helped to improve the original
manuscript.
N.H. and M.G.L. acknowledge the support of the BK21 program of the
Korean Government. This work was supported in part by a grant
(R01-2007-000-20336-0) from the Basic Research Program of the
Korea Science and Engineering Foundation.

\clearpage

\begin{deluxetable}{rcccccccccccccccc}
\tabletypesize{\scriptsize}
\setlength{\tabcolsep}{0.005in}
\rotate
\tablecaption{A Catalog of Class 1 Bright Star Clusters in M51\label{gcat}\tablenotemark{a}}
\tablewidth{0pt}

\tablehead{ \colhead{ID} & \colhead{RA (J2000)} & \colhead{Dec
(J2000)} & \colhead{$V$\tablenotemark{b}} & \colhead{err($V$)} &
\colhead{$(B-V)$\tablenotemark{b}} & \colhead{err$(B-V)$} &
\colhead{$(V-I)$\tablenotemark{b}} & \colhead{err$(V-I)$} &
\colhead{flag} & \colhead{stellarity} & \colhead{FWHM} &
\colhead{R$_{\rm eff}$} & \colhead{$(\chi^2/\chi^2_0)$} &
\colhead{ellipticity} &
\colhead{$V_{apc}$\tablenotemark{b}} & \colhead{remarks} \\
 & \colhead{[deg]} & \colhead{[deg]} & \colhead{[mag]} &  & \colhead{[mag]} &  & \colhead{[mag]} &  &
 &  & \colhead{[arcsec]} &  \colhead{[pc]} & & & \colhead{[mag]} &  }
\startdata
    459 & 202.4433289 &  47.1328163 &  22.146 &   0.015 &   0.873 &   0.027 &   1.223 &   0.018  &    0 & 0.03 &  0.197 &   4.99 &  0.023 &  0.078 &  22.002 &   ... \\
    498 & 202.4610443 &  47.1324654 &  21.258 &   0.010 &   0.328 &   0.015 &   0.794 &   0.013  &    0 & 0.03 &  0.278 &   3.43 &  0.259 &  0.081 &  21.258 &   ... \\
    543 & 202.4734344 &  47.1334114 &  22.967 &   0.024 &   0.833 &   0.041 &   1.193 &   0.028  &    0 & 0.00 &  0.371 &  11.89 &  0.053 &  0.151 &  22.614 &    FF \\
    707 & 202.4062195 &  47.1339684 &  22.023 &   0.014 &   0.234 &   0.021 &   0.524 &   0.020  &    0 & 0.03 &  0.199 &   4.53 &  0.153 &  0.341 &  21.847 &   ... \\
    875 & 202.4083099 &  47.1347275 &  22.173 &   0.016 &   0.132 &   0.022 &   0.363 &   0.022  &    0 & 0.03 &  0.222 &   5.06 &  0.099 &  0.404 &  21.908 &   ... \\
   1084 & 202.4343414 &  47.1359024 &  22.832 &   0.022 &   0.237 &   0.032 &   0.539 &   0.031  &    0 & 0.02 &  0.166 &   3.06 &  0.175 &  0.109 &  22.552 &   ... \\
   1249 & 202.5207825 &  47.1363945 &  21.969 &   0.014 &   0.771 &   0.024 &   1.158 &   0.017  &    0 & 0.07 &  0.151 &   2.55 &  0.057 &  0.049 &  21.936 &   ... \\
   1312 & 202.4645996 &  47.1362038 &  22.199 &   0.016 &   0.325 &   0.024 &   0.930 &   0.020  &    0 & 0.03 &  0.161 &   2.85 &  0.104 &  0.148 &  22.026 &   ... \\
   1366 & 202.4916687 &  47.1367531 &  22.033 &   0.015 &   0.621 &   0.024 &   1.002 &   0.018  &    0 & 0.04 &  0.146 &   2.23 &  0.103 &  0.076 &  21.791 &   ... \\
   1854 & 202.3926697 &  47.1383667 &  22.934 &   0.023 &   0.871 &   0.040 &   1.195 &   0.028  &    0 & 0.02 &  0.170 &   3.50 &  0.079 &  0.008 &  22.676 &   ... \\
\enddata
\tablenotetext{a}{The complete version of this table is in the
electronic edition of the Journal. The printed edition contains
only a sample.} \tablenotetext{b}{All photometric data in this
table are not given in the standard Johnson system but in the
$HST$ equivalents. See Section \ref{phot} for details.}
\end{deluxetable}

\clearpage

\begin{thebibliography}{}
\bibitem[Bastian et al. (2005a)]{bas05a} Bastian, N., Gieles, M., Lamers, H. J. G. L.
M., Scheepmaker, R. A., \& de Grijs, R. 2005a, \aap, 431, 905
\bibitem[Bastian et al. (2005b)]{bas05b} Bastian, N., Gieles, M., Efremov, Y., \& Lamers, H. J. G. L. M.
2005b, \aap, 443, 79
\bibitem[Bertin \&  Arnouts(1996)]{ber96}Bertin, E., \& Arnounts, S. 1996, \aaps, 117, 393
\bibitem[Mutchler et al. (2005)]{mut05} Mutchler, M. et al. 2005, BAAS, 37, 2
\bibitem[Mackey et al. (2006)]{mac06} Mackey, A. D. et al. 2006, \apj, 653, L105
\bibitem[Bik et al. (2003)]{bik03}Bik, A., Lamers, H. J. G. L. M., Bastian, N., Panagia, N., \& Romaniello, M. 2003, \aap, 397, 473
\bibitem[Brodie \& Larsen (2002)]{bro02}Brodie, J. P., \& Larsen, S. S. 2002, \aj, 124, 1410
\bibitem[Bruzual \& Charlot (2003)]{bc03}Bruzual, \& Charlot, 2003
\bibitem[Chandar et al. (2004)]{cha04}Chandar, R., Whitmore, B., \& Lee, M. G. 2004, \aj, 611, 220
\bibitem[de Vaucouleurs et al. (1991)]{dev91}de Vaucouleurs, G.  et al. 1991, Third Reference Catalog of Bright Galaxies (Berlin:Springer)
\bibitem[Feldmeier et al. (1997)]{fel97}Feldmeier, J. J., Ciardullo, R., \& Jacoby, G. H. 1997, \apj, 479, 231
\bibitem[Gieles et al. (2005)]{gie05} Gieles, M., Bastian, N., Lamers, H. J. G. L.
M., \& Mout, J. N. 2005, \aap, 441, 949
\bibitem[Gieles et al. (2006a)]{gie06a} Gieles, M., Larsen, S. S., Scheepmaker, R. A., Bastian, N., Haas, M. R.,
\& Lamers, H. J. G. L. M. 2006a, \aap, 446, L9
\bibitem[Gieles et al. (2006b)]{gie06b} Gieles, M., Larsen, S. S., Bastian, N., \& Stein, I. T.
2006b, \aap, 450, 129
\bibitem[Harris (1996)]{har96}Harris, W. 1996, \aj, 112, 1487
\bibitem[Huxor et al. (2005)]{hux05} Huxor, A. P. et al. 2005, \mnras, 360, 1007
\bibitem[Hwang et al. (2005)]{hwa05} Hwang, N., et al. 2005, IAU Colloquium, 198, Near Field Cosmology With Dwarf Elliptical Galaxies, Eds. H. Jerjen \& B. Bingeli (Cambridge: Cambridge Univ. Press), 257
\bibitem[Hwang \& Lee (2006)]{hwa06} Hwang, N., \& Lee, M. G. 2006, \apj, 638, L79
\bibitem[Hwang \& Lee (2007a)]{hwa07a} Hwang, N., \& Lee, M. G. 2007a, IAU Symposium, 241, Stellar Populations as Building Blocks of Galaxies,
Eds. A. Vazdekis, \& R. Peletier (Cambridge: Cambridge University Press), 451
\bibitem[Hwang \& Lee (2007b)]{hwa07b} Hwang, N., \& Lee, M. G. 2007b, in preparation.
\bibitem[Larsen(1999)]{lar99}Larsen, S. S. 1999, \aaps, 139, 393
\bibitem[Larsen (2002)]{lar02} Larsen, S. S. 2002, \aj, 124, 1393
\bibitem[Larsen \& Brodie (2000)]{larb00}Larsen, S. S., \& Brodie, J. P. 2000, \aj, 120, 2938
\bibitem[Lee (2003)]{lee03}Lee, M. G. 2003, Journal of Korea Astr. Soc., 36, 189
\bibitem[Lee et al. (2005)]{lee05}Lee, M. G., Chandar, R., \& Whitmore, B. W. 2005, \aj, 130, 2128
\bibitem[Lee (2006)]{lee06}Lee, M. G. 2006, Bulletin of Astr. Soc. India, 34, 99
\bibitem[Lee et al. (2007)]{lee07} Lee, J. H. et al. 2007, in preparation.
\bibitem[Peng et al.(2006)]{pen06} Peng, E. W., et al. 2006, \apj, 639, 838
\bibitem[Rand (1992)]{ran92a}Rand, R. J. 1992, \aj, 103, 815
\bibitem[Rots et al. (1990)]{rot90}Rots, A. H., Bosma, A., van der Hulst, J. M.,  Athanassoula, E., \& Crane, P. C. 1990, \aj, 100, 387
\bibitem[Salo \& Laurikainen (2000a)]{sal00a}Salo, H., \& Laurikainen, E. 2000a, \mnras, 319, 377
\bibitem[Salo \& Laurikainen (2000b)]{sal00b}Salo, H., \& Laurikainen, E. 2000b, \mnras, 319, 393
\bibitem[Scheepmaker et al. (2006)]{sch06}
Scheepmaker, R. A., Gieles, M., Haas, M. R., Bastian, N., Larsen,
S. S., \& Lamers, H. J. G. L. M. 2006, astro-ph/0605022
\bibitem[Scheepmaker et al. (2007)]{sch07}
Scheepmaker, R. A., Haas, M. R., Gieles, M., Bastian, N., Larsen,
S. S., \& Lamers, H. J. G. L. M. 2007, \aap, 469, 925
\bibitem[Schlegel et al. (1998)]{sch98} Schlegel, D. J., Finkbeiner, D. P., \& Davis, M. 1998, \apj, 500, 525
\bibitem[Schuster et al. (2007)]{schu07} Schuster, K. F., Kramer,
C., Hitschfeld, M., Garcia-Burillo, S., \& Mookerjea, B. 2007,
\aap, 461, 143
\bibitem[Sirianni et al. (2005)]{sir05}Sirianni, M., et al. 2005, \pasp, 117, 1049
\bibitem[Toomre \& Toomre (1972)]{too72}Toomre, R. P. J., \& Toomre, J. 1972, \apj, 178, 623
\bibitem[van den Bergh \& Mackey (2004)]{vdb04} van den Bergh, S., \& Mackey, A. D. 2004, \mnras, 354, 713
\bibitem[Whitmore et al. (1999)]{whi99}Whitmore, B. C., Zhang, Q., Leitherer, C., Fall, S. M., Schweizer, F., \& Miller, B. 1999, \aj, 118, 1551
\end{thebibliography}
\end{document}